\documentclass[12pt]{article}
\usepackage{amsfonts,amssymb,epsfig,amsmath,mathtools}
\usepackage{color}

\usepackage{mathrsfs}
\usepackage{graphicx}
\usepackage{subcaption}
\renewcommand{\baselinestretch}{1.2}

\begin{document}

\makeatletter \@addtoreset{equation}{section} \makeatother
\renewcommand{\theequation}{\thesection.\arabic{equation}}
\renewcommand{\thefootnote}{\alph{footnote}}

\begin{titlepage}

\begin{center}
\hfill {\tt KIAS-P21054}\\
\hfill {\tt SNUTP21-002}\\

\vspace{2cm}

{\Large\bf Exact QFT duals of AdS black holes}

\vspace{2cm}

\renewcommand{\thefootnote}{\alph{footnote}}

{\large Sunjin Choi$^{1}$, Saebyeok Jeong$^2$, Seok Kim$^3$ and
Eunwoo Lee$^3$}

\vspace{0.3cm}

\textit{$^1$School of Physics, Korea Institute for Advanced Study,\\
85 Hoegiro, Dongdaemun-gu, Seoul 02455, Korea.}\\

\vspace{0.2cm}

\textit{$^2$New High Energy Theory Center, Rutgers University,\\
136 Frelinghuysen Road, Piscataway, New Jersey 08854-8019, USA.}\\

\vspace{0.2cm}

\textit{$^3$Department of Physics and Astronomy \& Center for
Theoretical Physics,\\
Seoul National University, 1 Gwanak-ro, Seoul 08826, Korea.}\\

\vspace{0.7cm}

E-mails: {\tt sunjinchoi@kias.re.kr,
saebyeok.jeong@physics.rutgers.edu, seokkimseok@gmail.com,
eunwoo42@snu.ac.kr}

\end{center}

\vspace{1cm}

\begin{abstract}

We construct large $N$ saddle points of the matrix model for the
$\mathcal{N}=4$ Yang-Mills index dual to the BPS black holes in
$AdS_5\times S^5$, in two different setups.
When the two complex chemical potentials for the angular momenta are
collinear, we find linear eigenvalue distributions which solve the
large $N$ saddle point equation. When the chemical potentials are not collinear,
we find novel solutions given by areal eigenvalue distributions after
slightly reformulating the saddle point problem.
We also construct a class of multi-cut saddle points, showing that
they sometimes admit nontrivial filling fractions.
As a byproduct, we find that the Bethe ansatz equation
emerges from our saddle point equation.

\end{abstract}

\end{titlepage}

\renewcommand{\thefootnote}{\arabic{footnote}}

\setcounter{footnote}{0}

\renewcommand{\baselinestretch}{1}

\tableofcontents

\renewcommand{\baselinestretch}{1.2}

\section{Introduction}

We want to better understand the microscopic black hole physics in AdS/CFT
\cite{Witten:1998zw,Sundborg:1999ue,Aharony:2003sx}.
For quantitative studies, BPS black holes
\cite{Gutowski:2004ez,Gutowski:2004yv,Chong:2005da,Kunduri:2006ek} are ideal
objects. In 2005, \cite{Kinney:2005ej,Romelsberger:2005eg} defined  the indices of
4d SCFTs and explored their large $N$ behaviors. It took some time to
understand how to see the black holes from this index
\cite{Cabo-Bizet:2018ehj,Choi:2018hmj,Benini:2018ywd}.
In 4d $\mathcal{N}=4$ Yang-Mills theory, we study the black holes in
$AdS_5\times S^5$. For this problem, one should
study an apparently complicated large $N$ matrix integral
(see section 2 for more precise statements),
\begin{equation}\label{matrix-integral}
  Z(\delta_I,\sigma,\tau)\sim\frac{1}{N!}\prod_{a=1}^N
  \int_{-\frac{1}{2}}^{\frac{1}{2}} du_a
  \cdot\prod_{a\neq b}\frac{\prod_{I=1}^3\Gamma(\delta_I+u_{ab},\sigma,\tau)}
  {\Gamma(u_{ab},\sigma,\tau)}\ \ ,\ \ \ u_{ab}\equiv u_a-u_b\ ,
\end{equation}
where $\sum_I\delta_I=\sigma+\tau$ (mod $\mathbb{Z}$).
$\Gamma(z,\sigma,\tau)$ is the elliptic gamma function.
In this paper, we construct its exact large $N$ saddle points.
Here we sketch their important features.

The integration contours of the $N$ eigenvalues $u_a$ are real
circles, $u_a\sim u_a+1$. One has to set the chemical potentials $\delta_I,\sigma,\tau$ complex. (See \cite{Agarwal:2020zwm,Choi:2021lbk} to
understand why.) The saddle points for $u_a$'s are also
complex. At our saddle point, the eigenvalues are typically distributed
\textit{areally}, in a 2 dimensional region on the complex $u$ plane.
We have not heard of areal distributions in holomorphic matrix
models: one finds linear eigenvalue distributions called `cuts.'
Our claim is true only after carefully setting up the saddle point problem
as we explain shortly.

\begin{figure}[!t]
\centering
\begin{subfigure} [b]{\textwidth}
\includegraphics[width=0.3\textwidth]{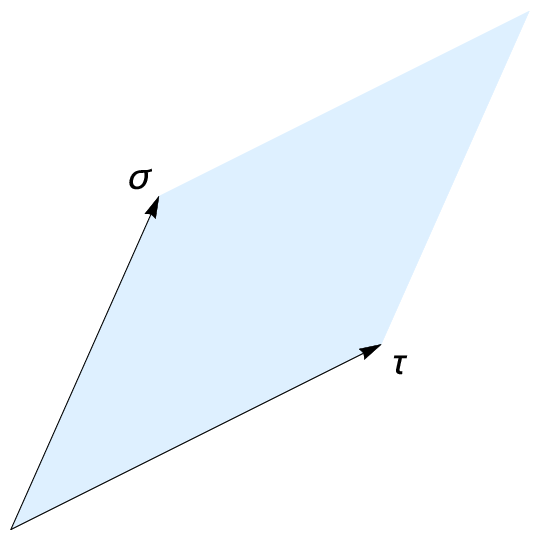}
\includegraphics[width=0.33\textwidth]{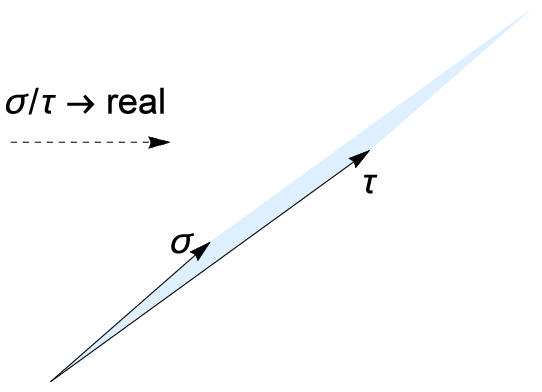}
\includegraphics[width=0.33\textwidth]{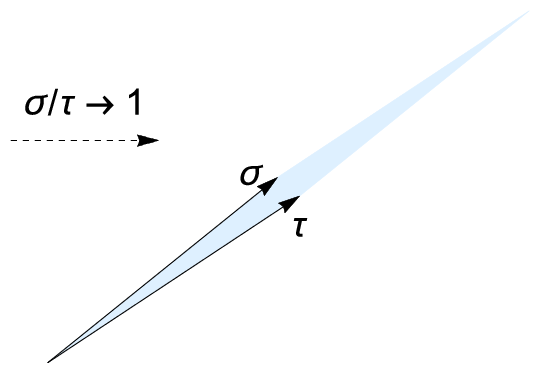}
\subcaption{Uniform parallelogram distribution on the complex $u$ plane
(left); linear degeneration at $\frac{\sigma}{\tau}\rightarrow$ real
$\neq 1$ (middle);
linear degeneration at $\frac{\sigma}{\tau}\rightarrow 1$ (right).}
\end{subfigure}
\quad
\begin{subfigure} [b]{0.45\textwidth}
\includegraphics[width=\textwidth]{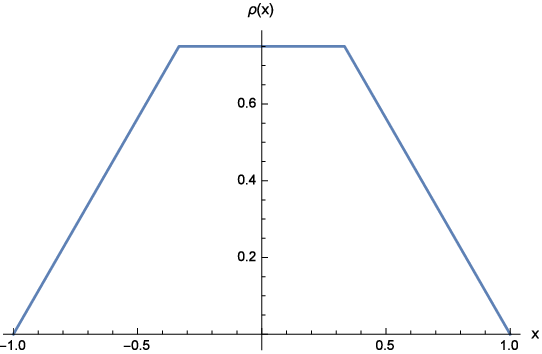}
\subcaption{Eigenvalue density $\rho(x)$ at
$\frac{\sigma}{\tau}=\frac{1}{2}$ (middle case of Fig.1(a)).
}
\end{subfigure}
\quad\hspace{0.8cm}
\begin{subfigure} [b]{0.45\textwidth}
\includegraphics[width=\textwidth]{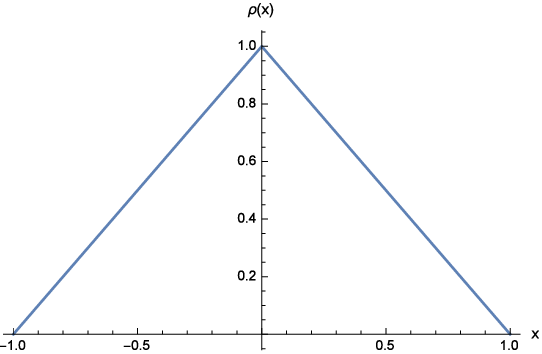}
\subcaption{Eigenvalue density $\rho(x)$ at $\sigma=\tau$
(right case of Fig.1(a)).}
\end{subfigure}
\caption{Illustrating the uniform parallelogram distribution,
and how it degenerates to linear cuts. In Figs. (b) and (c), $x\in(-1,1)$ parametrizes the linear cut.}
\label{parallel-cartoon}
\end{figure}

We explain our basic saddle point which corresponds to
the black hole solutions of \cite{Kunduri:2006ek}.
The saddle point depends on two parameters $\sigma$, $\tau$ of
the elliptic gamma function, related to the chemical potentials
for the spatial rotations.
$u_a$ are distributed uniformly inside a parallelogram, whose two
edge vectors are given by the complex numbers $\sigma$, $\tau$.
See Fig. 1(a). This is typically a 2d distribution when
$\sigma,\tau$ are not collinear.
When $\sigma$, $\tau$ are collinear, the parallelogram degenerates
to a line. In this limit, the density function on
the line is given by a trapezoid: see Fig. 1(b). When $\sigma=\tau$,
corresponding to equal angular momenta $J_1=J_2$, the linear
density function is triangular: see Fig. 1(c). These linear
density functions were discovered recently in the small \cite{Choi:2021lbk} and
large \cite{GonzalezLezcano:2020yeb} black hole limits, using different approaches.
Our parallelogram ansatz was inferred by guessing how to naturally
get the trapezoid limit.

For general non-collinear $\sigma$ and $\tau$,
this distribution solves the following saddle point equation.
We first note the integral identity (whose proof is reviewed in section 2)
\begin{equation}\label{haar-identity}
  \int_{-\frac{1}{2}}^{\frac{1}{2}} d^Nu \cdot
  \frac{1}{N!}\prod_{a<b}(1-e^{2\pi i\kappa u_{ab}})(1-e^{-2\pi i\kappa u_{ab}})
  \cdot f(u)=
  \int_{-\frac{1}{2}}^{\frac{1}{2}} d^Nu\cdot
  \prod_{a<b}(1-e^{2\pi i\kappa u_{ab}})\cdot f(u)\ ,
\end{equation}
which holds for any constant $\kappa$ and any permutation-invariant function $f(u)$.
For $\kappa=1$, the factor $\frac{1}{N!}\prod_{a<b}(1-e^{2\pi iu_{ab}})
(1-e^{-2\pi iu_{ab}})$ is the Haar measure of the unitary matrix integral and
this identity leads to the so-called Molien-Weyl formula \cite{Hanany:2008sb}.
In our problem, we shall write (\ref{matrix-integral}) in a way that it looks like
the left hand side of (\ref{haar-identity}) at $\kappa=\frac{1}{\sigma}$
or $\frac{1}{\tau}$. One can set up the saddle point problem
using either the left or right hand sides: integrals are the same,
but the saddle point problems are slightly different. Our parallelogram
solves the saddle point problem defined using the right hand side.
See section 2 for the precise setup of this problem.\footnote{It is not
uncommon to rewrite integrals using the symmetry of the integrand
to find simplified saddle point solutions. For instance,
this idea was used in the $S^3$ partition functions of large $N$ quiver
SCFTs \cite{Gang:2011jj}.}
As for the saddle point equation of the left hand side, presumably more
complicated saddles exist with same physical properties, but we have nothing
concrete to say at generic non-collinear $\sigma,\tau$.

In the collinear limit, $\frac{\sigma}{\tau}\rightarrow$ real,
the parallelogram degenerates to
linear distributions. In this case, our ansatz solves both saddle point
equations, defined using the left and right hand sides of (\ref{haar-identity}).
We explain how the saddle point equation of the left hand side is solved
in section 3. Furthermore, at $\sigma=\tau$, we show that the saddle point
equation of the left hand side of (\ref{haar-identity}) is related
to the Bethe ansatz equation for this index \cite{Closset:2017bse,Benini:2018mlo},
once employing our ansatz.

Our basic saddle points account for the physics of the
4-parameter BPS black hole solutions in $AdS_5\times S^5$
\cite{Kunduri:2006ek}, carrying three
angular momenta in $S^5$ and two in $AdS_5$.

We also find multi-cut solutions. For technical reasons, we only discuss
the case with collinear $\sigma,\tau$. We find
$K$-cut saddles given by $K$ parallelograms degenerated as Fig.
\ref{parallel-cartoon}. When $\sigma=\tau$, they are related to some Bethe
roots of \cite{Hong:2018viz}, labeled by two integers $K$, $r$.
Our $K$-cut solutions are related to the Bethe roots at $r=0$, and sometimes
admit generalization with unequal filling fractions.
The solutions at $r\neq 0$ may be
generated using the methods of appendix A.
(At $\sigma\neq \tau$, the extra parameter $r$
is further refined to two integers $r,s$.)

This paper is organized as follows. In section 2, we
derive the parallelogram saddle point.
In section 3, we discuss alternative derivation
of the saddles at collinear $\sigma,\tau$. Section 3.1 explains a relation
to the Bethe ansatz at $\sigma=\tau$. In section 4, we derive
the $K\geq 2$ cut saddle points. Section 5 concludes
with comments and future directions. Appendix A
discusses the $(r,s)$-refinements.

\section{Basic setup and the areal saddle points}

To highlight the key ideas, we first discuss general
matrix models satisfying certain conditions and derive the saddle points
given by areal distribution. We shall later show that the
matrix model for the $\mathcal{N}=4$ index belongs to this class.
Consider the $U(N)$ unitary matrix $U$ diagonalized to
$U={\rm diag}(e^{2\pi iu_1},\cdots,e^{2\pi iu_N})$. We consider
the following `matrix integral'
\begin{equation}\label{toy-full-haar}
  Z=\frac{1}{N!}\prod_{a=1}^N\int_{-\frac{1}{2}}^{\frac{1}{2}} du_a\cdot
  \prod_{a\neq b}(1-e^{\frac{2\pi iu_{ab}}{\tau}})\cdot
  \exp\left[-\sum_{a\neq b}(V_\sigma(u_{ab})+V_\tau(u_{ab}))\right]\ ,
\end{equation}
where $u_{ab}\equiv u_a-u_b$. The potential consists of two
terms, $V_\sigma(u)$ and $V_\tau(u)$, which we take to be holomorphic and
periodic in two different directions given by the complex numbers
$\sigma,\tau$:
\begin{equation}
  V_\sigma(u+\sigma)=V_\sigma(u)\ ,\ \
  V_\tau(u+\tau)=V_\tau(u)\ .
\end{equation}
The functions $V_\sigma,V_\tau$ are assumed to contain no singularities
such as poles or branch points in the region of $u_a$'s to be specified below.

We would like to apply (\ref{haar-identity}) at $\kappa=\frac{1}{\tau}$
to this integral. Before this, we review how to prove this identity. First note
that
\begin{equation}\label{haar-vandermonde}
  \frac{1}{N!}\prod_{a<b}(1-e^{x_a-x_b})(1-e^{x_b-x_a})=
  \frac{1}{N!}\prod_{a<b}(e^{x_b}-e^{x_a})(e^{-x_b}-e^{-x_a})\equiv
  \frac{1}{N!}\prod_{a<b}(\lambda_b-\lambda_a)(\lambda_b^{-1}-\lambda_a^{-1})\ ,
\end{equation}
where $\lambda_a\equiv e^{x_a}\equiv e^{2\pi i\kappa u_a}$.
We then recall the following formula for the Vandermonde matrix:
\begin{equation}
  \prod_{a<b}(\lambda_b-\lambda_a)=\det(\lambda_a^{b-1})=
  \sum_{\rho\in S_N}(-1)^{\epsilon(\rho)}\prod_{a=1}^N(\lambda_{\rho(a)})^{a-1}\ ,
\end{equation}
and similar formula for $\prod_{a<b}(\lambda_b^{-1}-\lambda_a^{-1})$.
$S_N$ is the permutation group, and $\epsilon(\rho)$ is the signature
of its element $\rho$. Applying these formulae, (\ref{haar-vandermonde}) can
be written as
\begin{equation}
  \frac{1}{N!}\sum_{\rho,\sigma\in S_N}
  (-1)^{\epsilon(\rho)}(-1)^{\epsilon(\sigma)}
  \prod_{a=1}^N(\lambda_{\rho(a)})^{a-1}(\lambda_{\sigma(a)}^{-1})^{a-1}\ .
\end{equation}
At fixed $\sigma$, one can relabel $\rho$ as
$\rho\cdot \sigma$ and write
\begin{eqnarray}\label{molien-identity}
  \frac{1}{N!}\sum_\sigma\sum_{\rho}(-1)^{\epsilon(\rho)}\prod_{a=1}^N
  (\lambda_{\rho(\sigma(a))})^{a-1}(\lambda_{\sigma(a)}^{-1})^{a-1}
  &=&\frac{1}{N!}\sum_\sigma
  \left[\prod_{a}(\lambda_{\sigma(a)})^{-(a-1)}\right]\cdot
  \left[\prod_{a<b}(\lambda_{\sigma(b)}-\lambda_{\sigma(a)})\right]\nonumber\\
  &=&\frac{1}{N!}\sum_\sigma\prod_{a<b}(1-\lambda_{\sigma(a)}
  \lambda_{\sigma(b)}^{-1})\ .
\end{eqnarray}
Now consider the integral (\ref{haar-identity}) with the Haar measure like
factor rewritten as the second line of (\ref{molien-identity}).
This is given by the sum of $N!$ integrals labeled by $\sigma\in S_N$,
divided by $N!$. Since $f(u)$ and the integral domain
$-\frac{1}{2}<u_a<\frac{1}{2}$ are all invariant under the permutations,
all $N!$ integrals yield same values. Therefore one proves the identity
(\ref{haar-identity}).

So our matrix integral can be rewritten as
\begin{equation}\label{toy-half-haar}
  Z=\prod_{a=1}^N\int_{-\frac{1}{2}}^{\frac{1}{2}} du_a\cdot
  \prod_{a<b}(1-e^{\frac{2\pi iu_{ab}}{\tau}})\cdot
  \exp\left[-\sum_{a\neq b}(V_\sigma(u_{ab})+V_\tau(u_{ab}))\right]\ .
\end{equation}
The half-Haar measure like factor can also be exponentiated and contribute to
the potential as
\begin{equation}\label{potential-half-haar}
  -V\leftarrow \sum_{a<b}\log(1-e^{\frac{2\pi iu_{ab}}{\tau}})\ .
\end{equation}
One can regard it as modifying the
$\tau$-periodic potential $V_\tau$.
We write $Z$ as
\begin{equation}\label{integral-half-haar}
  Z=\prod_{a=1}^N\int_{-\frac{1}{2}}^{\frac{1}{2}} du_a\cdot
  \exp\left[-\sum_{a\neq b}(V^{({\rm sgn}(a-b))}_\tau(u_{ab})
  +V_\sigma(u_{ab}))\right]\ ,
\end{equation}
where
\begin{equation}\label{V-pm}
  V^{(+)}_\tau(u)\equiv V_\tau(u)\ ,\ \
  V^{(-)}_\tau(u)\equiv V_\tau(u)-\log(1-e^{\frac{2\pi iu}{\tau}})\ .
\end{equation}
If the extra term in $V^{(-)}$ does not contain branch points in the region of our
interest, $V^{(\pm)}$ will also be holomorphic.

We shall find a large $N$ saddle point of  (\ref{integral-half-haar}),
and deform the integral contour to reach this saddle point.
The integral domain $u_a\in(-\frac{1}{2},\frac{1}{2})$ has a boundary.
(In our Yang-Mills matrix model,
the original problem is not on an interval due to the $u_a\sim u_a+1$ periods,
but the integrand after applying (\ref{haar-identity}) is not periodic.)
Knowledgeable readers may be concerned that, when the integral contour has a boundary,
the saddle point problem is well-posed only if the integrand vanishes at
the boundary. (More formally, this is the condition for the Picard-Lefschetz theory
of saddle point approximation to be applicable.)
We can rephrase the saddle point approximation of our integral in the standard
fashion, along a noncompact contour with the integrand vanishing at infinity,
as follows. Relabeling the integral variables as
\begin{equation}\label{relabel}
  u_a=\frac{1}{2}\tanh x_a\ ,
\end{equation}
the domain $(-\frac{1}{2},\frac{1}{2})$ for $u_a$ maps to
the real axis $(-\infty,\infty)$ for $x_a$. The saddle point problem
can then be phrased in terms of $x_a$, in which case one finds the extra
contribution
\begin{equation}\label{potential-jacobian}
  -V(x)\leftarrow \sum_{a=1}^N\log\left(\frac{1}{2}{\rm sech}^2 x_a\right)
\end{equation}
added to the eigenvalue potential due to the Jacobian factor.
Taking derivative of the net potential with $x_a$,
the force from the original potential is at $N^1$ order, while that from
the added potential (\ref{potential-jacobian}) is at the subleading $N^0$ order
and is thus negligible. (The Jacobian factor may well matter when computing
subleading corrections in $1/N$.) Therefore, even if one considers a holomorphic
integral along compact intervals, the saddle points of the original potential
ignoring (\ref{potential-jacobian}) are relevant for the
large $N$ saddle point approximation.

Since the additional potential (\ref{potential-jacobian})
induced by relabelling $u_a\rightarrow x_a$ does not
affect the leading large $N$ saddle points, we keep working with $u_a$.
In the large $N$ limit, consider the uniform distribution of $N$ eigenvalues
on a parallelogram in the $u$ space. In the continuum description, the
eigenvalues label $a$ can be replaced by a tuple of continuous parameters
$a\rightarrow (x,y)$ where $-\frac{1}{2}<x,y<\frac{1}{2}$.
Our uniform parallelogram ansatz is given by
\begin{equation}\label{toy-parallelogram}
  u(x,y)=x\sigma+y\tau\ \ \ ,\ \ \ \rho(x,y)=1\ .
\end{equation}
$\rho(x,y)$ is the uniform areal density function satisfying $\int dxdy\rho(x,y)=1$.
Since the integrand of (\ref{toy-half-haar}) breaks Weyl invariance,
we order $u_a$'s as follows. Here we assume ${\rm Im}(\frac{\sigma}{\tau})>0$, and
order the $N$ eigenvalues in a way that $x_a>x_b$
if $a<b$. Then (\ref{potential-half-haar}) has no branch points
in the region $u_a=x_a\sigma+y_a\tau$, $x_a,y_a\in(-\frac{1}{2},\frac{1}{2})$
satisfying $x_a>x_b$ if $a<b$, since
\begin{equation}
  \left|e^{\frac{2\pi iu_{ab}}{\tau}}\right|=
  \left|e^{\frac{2\pi i\sigma x_{ab}}{\tau}}\right|<1
\end{equation}
for $a<b$. This means that we can stay in the principal branch of log
for the configuration (\ref{toy-parallelogram}) and its small deformations,
implying that $V_\sigma,V_\tau^{(\pm)}$ is holomorphic in this
region.\footnote{One may think that the new $V_\tau$
defined piecewise by $V_\tau^{(\pm)}(x\sigma+y\tau)$ has a step function singularity
at $x=0$. But in the fine-grained discrete picture,
we can assume that no two eigenvalues have precisely same values of $x$,
with minimal differences at order $N^{-\frac{1}{2}}$.}
We now show that (\ref{toy-parallelogram}) is a large $N$ saddle point
of the integral (\ref{integral-half-haar}).
We consider the following force acting on the eigenvalue $u_a$:
\begin{equation}
  \sum_{b(\neq a)}\left[\frac{\partial}{\partial u_a}
  V_\sigma(u_{ab})+\frac{\partial}{\partial u_a}
  V_\tau^{{\rm sgn}(a-b)}(u_{ab})\right]=
  -\sum_{b(\neq a)}\left[\frac{1}{\sigma}
  \frac{\partial}{\partial x_b}V_\sigma(u_{ab})
  +\frac{1}{\tau}\frac{\partial}{\partial y_b}
  V_\tau^{{\rm sgn}(a-b)}(u_{ab})\right]\ .\nonumber
\end{equation}
We used $\frac{\partial}{\partial u_a}\sim-\frac{\partial}{\partial u_b}$
and also the fact that $u_b=x_b\sigma+y_b\tau$ derivatives
can be replaced by either $\frac{1}{\sigma}\partial_{x_b}$ or
$\frac{1}{\tau}\partial_{y_b}$. In the large $N$ continuum limit,
the sum over $b$ is replaced by an integral over $x_b$, $y_b$ with the
areal eigenvalue density $\rho(x,y)=1$:
\begin{equation}
  -N\int_0^1 dx_2 dy_2\left[\frac{1}{\sigma}\partial_{x_2}V_\sigma(u_{12})
  +\frac{1}{\tau}\partial_{y_2}V_\tau^{{\rm sgn}(x_2-x_1)}(u_{12})\right]\ .
\end{equation}
Both terms in the integrand separately integrate to zero. For the first term,
one finds
\begin{equation}
  \int_0^1 dx_2 \partial_{x_2}V_\sigma(u_{12})=
  V_\sigma(u_1-\tau y_2-\sigma)-V_\sigma(u_1-\tau y_2)=0
\end{equation}
because
$V_\sigma$ is periodic in $\sigma$ shift.
As for the second term, one integrates over $y_2$ first
and similarly finds zero, from the periodicity of $V_\tau^{{\rm sgn}(x_2-x_1)}$.
This statement is invalid if branch points are inside the integral domain.
They are absent partly by assumption on $V_\tau$, and also because
the integrand contains half-Haar-like measure only.
If one starts from the saddle point problem with (\ref{toy-full-haar}),
one indeed finds a nonzero force from the terms with branch points.

We found a class of matrix models
which admits the uniform parallelogram solution: (1) The eigenvalue
potential decomposes to two functions $V_\sigma$, $V_\tau$ which have their
respective periods; (2) The potentials do not have branch points in the
parallelogram domain (\ref{toy-parallelogram}); (3) The matrix integrand
is changed to contain half-Haar like measure, using the identity
(\ref{haar-identity}).

Now we show that the matrix model for the index of the $\mathcal{N}=4$ Yang-Mills
theory belongs to this class, which will be proving that our parallelogram
distribution is a saddle point.
The index is defined by the trace over the Hilbert space
of the radially quantized CFT \cite{Romelsberger:2005eg,Kinney:2005ej}:
\begin{equation}\label{index-definition}
  Z(\Delta_I,\omega_i)\equiv{\rm Tr}
  \left[(-1)^Fe^{-\sum_{I=1}^3Q_I\Delta_I}e^{-\sum_{i=1}^2 J_i\omega_i}\right]
\end{equation}
where $Q_I,J_i$ are $U(1)^3\subset SO(6)$ R-charges and
$U(1)^2\subset SO(4)$ angular momenta.
$Q_I,J_i$ are half-integrally quantized for spinors. $Z$ is
defined on a 4-parameter space of chemical potentials,
$\sum_{I=1}^3\Delta_I-\sum_{i=1}^2\omega_i=0$, apparently mod $4\pi i\mathbb{Z}$
at this moment from the definition (\ref{index-definition}). The chemical potentials
should satisfy ${\rm Re}(\Delta_I)>0$ and ${\rm Re}(\omega_i)>0$ for the
trace to be well-defined.
The matrix integral for the index is given by
(\ref{matrix-integral}), which we repeat here:
\begin{equation}\label{matrix-integral-repeat}
  Z(\delta_I,\sigma,\tau)=\frac{1}{N!}\prod_{a=1}^N \int_{-\frac{1}{2}}^{\frac{1}{2}}
  du_a\cdot\prod_{a\neq b}\frac{\prod_{I=1}^3\Gamma(\delta_I+u_{ab},\sigma,\tau)}
  {\Gamma(u_{ab},\sigma,\tau)}\cdot
  \left[U(1)^N\ \textrm{part}\right]\ .
\end{equation}
The `$U(1)^N$ part' is the contribution from $N$ diagonal matrix
components of fields, which is independent of the integral variable
$u_a$ and makes an $\mathcal{O}(N^1)$ contribution to the free energy.
As we are interested in the leading (nonzero) free energy of order $N^2$,
we shall not distinguish the two expressions (\ref{matrix-integral}) and
(\ref{matrix-integral-repeat}) in this paper.
The parameters $\delta_I$, $\sigma$, $\tau$ are defined by
\begin{equation}
  \Delta_I=-2\pi i\delta_I\ ,\ \ \omega_1=-2\pi i\sigma\ ,\ \
  \omega_2=-2\pi i\tau\ .
\end{equation}
The elliptic gamma function is defined by
\begin{equation}
  \Gamma(z,\sigma,\tau)\equiv\prod_{m,n=0}^\infty
  \frac{1-e^{-2\pi iz}e^{2\pi i((m+1)\sigma+(n+1)\tau)}}
  {1-e^{2\pi iz}e^{2\pi i(m\sigma+n\tau)}}\ .
\end{equation}
It satisfies
$\Gamma(z,\sigma,\tau)=\Gamma(z+1,\sigma,\tau)=\Gamma(z,\sigma+1,\tau)=
\Gamma(z,\sigma,\tau+1)$. (Other properties of $\Gamma$
will be presented below when necessary.)
So the chemical potentials $\delta_I,\sigma,\tau$ all have
period $1$ in the expression (\ref{matrix-integral-repeat}). They
define an index on the following 4-parameter space:
\begin{equation}\label{4-para}
  \sum_{I=1}^3\delta_I=\sigma+\tau\ \mod \mathbb{Z}\ .
\end{equation}
They should further satisfy ${\rm Im}(\delta_I)>0$,
${\rm Im}(\sigma)>0$, ${\rm Im}(\tau)>0$.

We first set a convenient parametrization of $\delta_I$'s,
for given $\sigma,\tau$. We set $\delta_I=-a_I+b_I(\sigma+\tau)$ with
real coefficients $a_I,b_I$. From (\ref{4-para}), the coefficients
should satisfy $a_1+a_2+a_3\in \mathbb{Z}$ and
$b_1+b_2+b_3=1$. We shall first fix the ranges
of $a_I,b_I$. From ${\rm Im}(\delta_I)>0$
and $\delta_1+\delta_2+\delta_3=\sigma+\tau$ mod $\mathbb{Z}$,
one finds $0<{\rm Im}(\delta_I)<{\rm Im}(\sigma+\tau)$. So
all $b_I$'s are in the range $b_I\in(0,1)$.
The ranges of $a_I$ are fixed partly by convention, using
the periodicities of $\delta_I$.
Making shifts $\delta_I\rightarrow \delta_I+n_I$ with integral
$n_I$'s, each $a_I$ can be put in certain interval of length $1$.
For instance, one can set $a_I\in(0,1)$ for all $I=1,2,3$.
With this choice, one finds that $0<a_1+a_2+a_3<3$, so we
should take either $a_1+a_2+a_3=1$ or $2$.\footnote{Here and below,
we shall be loose about the boundary values, e.g. not sharply
distinguishing $a_I\in(0,1)$ and $a_I\in[0,1]$. But of course the latter
is correct and there are two more isolated choices $a_1+a_2+a_3=0$ or $3$.
These two cases are equivalent, corresponding to all $a_I$ being $0$.
But this point is again equivalent to a special point of (\ref{delta-classify}),
e.g. the upper case with $a_1=a_2=0$, $a_3=1$.}
In the former case, we have $\delta_1+\delta_2+\delta_3-\sigma-\tau=-1$
with $a_I\in(0,1)$. In the latter case, it is more
convenient to redefine $a_I$'s by shifting all of them
by $-1$, so that $a_I\in(-1,0)$ and $a_1+a_2+a_3=-1$.
In this convention, we take $\delta_1+\delta_2+\delta_3-\sigma-\tau=+1$
with $a_I\in(-1,0)$. To summarize, possible $\delta_I$'s
belong to one of the following two cases:
\begin{equation}\label{delta-classify}
  \delta_1+\delta_2+\delta_3=\sigma+\tau\mp 1\ ,\ \
  \delta_I=-a_I+b_I(\sigma+\tau)\ ,\ \
  \pm a_I\in(0,1)\ ,\ \ b_I\in(0,1).
\end{equation}
In the unrefined index with all equal R-charges,
we set $b_I=\frac{1}{3}$ and $a_I=\pm\frac{1}{3}$ for all $I$,
respectively. The two cases define mutually complex conjugate
regions in the chemical potential space,
in which the real parts of all $\delta_I$, $\sigma$, $\tau$ are sign-flipped.
In other words, if the complex numbers $(\delta_I,\sigma,\tau)$ belong to
the upper case, $(-\delta_I^\ast,-\sigma^\ast,-\tau^\ast)$ belong to the
lower case.

Now consider the following two identities of $\Gamma(z,\sigma,\tau)$
\cite{felder,Gadde:2020bov}
\begin{eqnarray}\label{modular}
  \Gamma(z,\sigma,\tau)&=&
  %e^{-\pi iQ_+(z,\sigma,\tau)}
  %\frac{\Gamma(\frac{z}{\tau},-\frac{1}{\tau},\frac{\sigma}{\tau})}
  %{\Gamma(\frac{z-\tau}{\sigma},-\frac{1}{\sigma},-\frac{\tau}{\sigma})}=
  e^{-\pi iQ_+(z,\sigma,\tau)}
  \Gamma({\textstyle \frac{z}{\tau},-\frac{1}{\tau},\frac{\sigma}{\tau}})
  \Gamma({\textstyle \frac{-z-1}{\sigma},-\frac{1}{\sigma},-\frac{\tau}{\sigma}})
  \nonumber\\
  \hspace*{-1cm}
  \Gamma(z,\sigma,\tau)&=&e^{-\pi iQ_-(z,\sigma,\tau)}
  \Gamma({\textstyle{-\frac{z}{\sigma}},-\frac{1}{\sigma},-\frac{\tau}{\sigma}})
  \Gamma({\textstyle \frac{z-1}{\tau},-\frac{1}{\tau},\frac{\sigma}{\tau}})
\end{eqnarray}
where
\begin{equation}
  Q_\pm=\frac{z^3}{3\sigma\tau}-\frac{\sigma+\tau\mp 1}{2\sigma\tau}z^2
  +\frac{\sigma^2+\tau^2+3\sigma\tau\mp 3\sigma\mp 3\tau+1}{6\sigma\tau}z
  +\frac{1}{12}(\sigma+\tau\mp 1)\left(\frac{1}{\sigma}+\frac{1}{\tau}\mp 1\right)\ .
\end{equation}
These are part of the $SL(3,\mathbb{Z})$ transformation identities.
One identity can be obtained from another by complex-conjugating an
identity and regarding $(-z^\ast,-\tau^\ast,-\sigma^\ast)$ as
the new $(z,\sigma,\tau)$ parameters.
We shall use these identities to rewrite the integrands of the matrix
model at ${\rm Im}(\frac{\sigma}{\tau})>0$, in which case
all the modular parameters have positive imaginary parts.
For ${\rm Im}(\frac{\sigma}{\tau})<0$, we can use the identities with
the role of $\sigma,\tau$ flipped and proceed similarly.
The limit of collinear $\sigma$ and $\tau$, i.e.
${\rm Im}(\frac{\sigma}{\tau})\rightarrow 0$, is singular
with each $\Gamma$ function.
However, the collinear limit can be smoothly taken after all the
calculation is done. This limit will be discussed in section 3.

It is convenient to use the first/second line
of (\ref{modular}) for the two cases of (\ref{delta-classify}) with upper/lower signs,
respectively. The integral for the upper case of (\ref{delta-classify})
can be written as
\begin{equation}\label{YM-matrix-canonical}
  Z=\exp\left[-\frac{\pi iN^2\delta_1\delta_2\delta_3}{\sigma\tau}\right]
  \frac{1}{N!}\int\prod_{a=1}^N du_a\cdot
  \exp\left[-\sum_{a\neq b}\left(V_\sigma(u_{ab})+\tilde{V}_\tau(u_{ab})\right)\right]
\end{equation}
where
\begin{eqnarray}\label{potential-1}
  -V_\sigma(u)&\equiv&\frac{1}{2}\sum_{I=1}^3\log\Gamma({\textstyle
  -\frac{\delta_I+u+1}{\sigma},-\frac{1}{\sigma},-\frac{\tau}{\sigma}})
  -\frac{1}{2}\log\Gamma({\textstyle -\frac{u+1}{\sigma},-\frac{1}{\sigma},
  -\frac{\tau}{\sigma}})+\left(u\rightarrow-u\right)
  \nonumber\\
  -\tilde{V}_\tau(u)&\equiv&\frac{1}{2}\sum_{I=1}^3\log\Gamma({\textstyle
  \frac{\delta_I+u}{\tau},-\frac{1}{\tau},\frac{\sigma}{\tau}})
  -\frac{1}{2}\log\Gamma({\textstyle \frac{u}{\tau},-\frac{1}{\tau},
  \frac{\sigma}{\tau}})+\left(u\rightarrow-u\right)\ ,
\end{eqnarray}
and for the lower case of (\ref{delta-classify}) can be written as
\begin{equation}\label{YM-matrix-canonical-lower}
  Z=\exp\left[-\frac{\pi iN^2\delta_1\delta_2\delta_3}{\sigma\tau}\right]
  \frac{1}{N!}\int\prod_{a=1}^N du_a\cdot
  \exp\left[-\sum_{a\neq b}
  \left(\tilde{V}_\sigma(u_{ab})+V_\tau(u_{ab})\right)\right]
\end{equation}
where
\begin{eqnarray}\label{potential-2}
  -\tilde{V}_\sigma(u)&\equiv&\frac{1}{2}\sum_{I=1}^3\log\Gamma({\textstyle
  -\frac{\delta_I+u}{\sigma},-\frac{1}{\sigma},-\frac{\tau}{\sigma}})
  -\frac{1}{2}\log\Gamma({\textstyle -\frac{u}{\sigma},-\frac{1}{\sigma},
  -\frac{\tau}{\sigma}})+\left(u\rightarrow-u\right)
  \nonumber\\
  -V_\tau(u)&\equiv&\frac{1}{2}\sum_{I=1}^3\log\Gamma({\textstyle
  \frac{\delta_I+u-1}{\tau},-\frac{1}{\tau},\frac{\sigma}{\tau}})
  -\frac{1}{2}\log\Gamma({\textstyle \frac{u-1}{\tau},-\frac{1}{\tau},
  \frac{\sigma}{\tau}})+\left(u\rightarrow-u\right)\ .
\end{eqnarray}
Here we averaged over the contributions from positive/negative roots to have $V_{\sigma,\tau},\tilde{V}_{\sigma,\tau}$ to be even functions of $u$.
The prefactor $\exp\left[-\frac{\pi iN^2\delta_1\delta_2\delta_3}{\sigma\tau}\right]$
is obtained by collecting all four $e^{-\pi iQ_\pm}$ factors of the
identity. Namely, one obtains
\begin{eqnarray}
  &&\hspace*{-0.5cm}-\pi i\left[\sum_{I=1}^3Q_\pm(\delta_I+u)-Q_\pm(u)\right]
  +(u\rightarrow -u)=-\frac{2\pi i\delta_1\delta_2\delta_3}{\sigma\tau}
  -\frac{2\pi iu^2\Delta_\pm}{\sigma\tau}\\
  &&\hspace*{0.5cm}
  -\pi i\frac{2\Delta_\pm^3+3\Delta_\pm^2(\sigma+\tau\mp 1)
  +\Delta_\pm\left(\sigma^2\!+\!\tau^2\!+\!3\sigma\tau\mp 3(\sigma+\tau)\!+\!1
  \!-\!6(\delta_1\delta_2+\delta_2\delta_3+\delta_3\delta_1)\right)}{3\sigma\tau}
  \ ,\nonumber
\end{eqnarray}
where $\Delta_\pm\equiv\sum_{I=1}^3\delta_I-\sigma-\tau\pm 1$. Since
$\Delta_\pm=0$ in the two cases of (\ref{delta-classify}), respectively,
this factor simplifies to a constant
$-\frac{2\pi i\delta_1\delta_2\delta_3}{\sigma\tau}$
in both cases. Then we can take
$\frac{N^2-N}{2}\approx \frac{N^2}{2}$ such factors out of the integral to
obtain the prefactor of (\ref{YM-matrix-canonical}).

Since the analysis is completely the same in the two cases of (\ref{delta-classify}),
here we only discuss the case with upper signs.
Apart from the prefactor,
(\ref{YM-matrix-canonical}) takes the form of (\ref{toy-full-haar}),
as we explain now. First of all, the $\Gamma$ functions appearing in
$V_\sigma$ and $\tilde{V}_\tau$ are periodic in shifting $u$ by $\sigma$ and $\tau$, respectively. So if one does not cross the branch
cuts for $\log \Gamma$ as one parallel shifts $u$ by $\sigma$ and $\tau$,
one would find $V_\sigma(u+\sigma)=V_\sigma(u)$,
$\tilde{V}_\tau(u+\tau)=\tilde{V}_\tau(u)$.
We will show that $V_\sigma$ has no branch points in the parallelogram
region defined by $u_a=\sigma x_a+\tau y_a$
with $x_a,y_a\in(-\frac{1}{2},\frac{1}{2})$,
and that all log functions of the form $\log(1-x)$ can be defined by
Taylor expansion $-\sum_{n=1}^\infty\frac{x^n}{n}$.
On the other hand, $\tilde{V}_\tau$ will have branch points/cuts
from a Haar-measure-like factor which looks like the left hand side
of (\ref{haar-identity}) with $\kappa=\frac{1}{\tau}$. After one proves
these two assertions above, one can separate the dangerous Haar-measure-like
factor and define $V_\tau(u)$ by
\begin{equation}\label{V-tilde-V}
  e^{-\sum_{a\neq b}\tilde{V}_\tau(u_{ab})}=
  \prod_{a\neq b}(1-e^{\frac{2\pi iu_{ab}}{\tau}})
  \cdot e^{-\sum_{a\neq b}V_\tau(u_{ab})}\ ,
\end{equation}
and apply (\ref{haar-identity}) to derive a matrix model
of the form of (\ref{toy-half-haar}). Then from our general discussion
earlier, the parallelogram saddle point is trivially derived.

Now we only need to prove the key assertion, about the absence of
branch points for $V_\sigma$ and also for $V_\tau$
defined by (\ref{V-tilde-V}),
in the region $u_a=x_a\sigma+y_a\tau$, $u_b=x_b\sigma+y_b\tau$ with
$x_a,x_b,y_a,y_b\in(-\frac{1}{2},\frac{1}{2})$.
We shall prove this when certain conditions
for $\delta_I,\sigma,\tau$ are met. Namely, we shall be deriving
the `basic parallelogram saddle point' in certain region of the parameter space.
In appendix A, we derive more general saddle points,
virtually for any value of $\delta_I,\sigma,\tau$.

We first show that $\tilde{V}_\tau$ contains branch points
only from the Haar measure like factor
\begin{equation}\label{V-tau-exception}
  \prod_{a\neq b}(1-e^{\frac{2\pi iu_{ab}}{\tau}})\ .
\end{equation}
It suffices to show that
$\Gamma(\frac{\delta_I+u_{12}}{\tau},-\frac{1}{\tau},\frac{\sigma}{\tau})$
and $\Gamma(\frac{u_{12}}{\tau},-\frac{1}{\tau},\frac{\sigma}{\tau})$
has no zeros or poles when $u_1$, $u_2$ are both in the parallelogram,
except for the zeros from (\ref{V-tau-exception}).
We first consider the function
\begin{equation}\label{Gamma-tau}
  \Gamma({\textstyle
  \frac{\delta+u_{12}}{\tau},-\frac{1}{\tau},\frac{\sigma}{\tau}})
  =\prod_{m,n=0}^\infty
  \frac{1-e^{2\pi i(-\frac{\delta+u_{12}}{\tau}
  -\frac{m+1}{\tau}+\frac{(n+1)\sigma}{\tau})}}
  {1-e^{2\pi i(\frac{\delta+u_{12}}{\tau}-\frac{m}{\tau}+\frac{n\sigma}{\tau})}}\ ,
\end{equation}
where $\delta\equiv -a+b(\sigma+\tau)$ with $a,b\in(0,1)$ is either
$\delta_{1,2,3}$. If the exponential factors appearing in the denominator
always have their absolute values smaller than $1$ for any $m,n$,
the denominator will not have poles.
This is true if the following imaginary part is always positive:
\begin{equation}
  {\rm Im}\left[\frac{\delta+u_{12}}{\tau}-\frac{m}{\tau}+\frac{n\sigma}{\tau}\right]
  ={\rm Im}({\textstyle -\frac{1}{\tau}})\left[m+a+\mathcal{N}(n+b+x_{12})\right]
\end{equation}
where $\mathcal{N}\equiv \frac{{\rm Im}(\frac{\sigma}{\tau})}
{{\rm Im}(-\frac{1}{\tau})}$ is a positive number. One finds
\begin{equation}
  m+a+\mathcal{N}(n+b+x_{12})\geq a-\mathcal{N}(1-b)\ ,
\end{equation}
where the inequality is saturated at $m=n=0$ and $x_{12}=-1$.
Therefore, this quantity remains positive
when $\mathcal{N}<\frac{a}{1-b}$. This is one of the conditions that
we shall assume for the chemical potentials.
Now considering the numerator of (\ref{Gamma-tau}), we similarly check
if the imaginary part of the exponent is always positive. The relevant
imaginary part is
\begin{equation}
  {\rm Im}\left[\frac{-\delta-u_{12}}{\tau}-\frac{m+1}{\tau}
  +\frac{(n+1)\sigma}{\tau}\right]=
  {\rm Im}({\textstyle -\frac{1}{\tau}})\left[
  m+1-a+\mathcal{N}(n+1-b-x_{12})\right]\geq
  1-a-b\mathcal{N}\ ,
\end{equation}
where the inequality is saturated at $m=n=0$ and $x_{12}=1$.
So we find that the imaginary part is positive
when $\mathcal{N}<\frac{1-a}{b}$, which we also assume.
So for
\begin{equation}\label{condition-V-tau}
  \mathcal{N}\equiv\frac{{\rm Im}(\frac{\sigma}{\tau})}{{\rm Im}(-\frac{1}{\tau})}
  <\min\left[\frac{a_I}{1-b_I},\frac{1-a_I}{b_I}\right]\ ,
\end{equation}
$\log \Gamma(\frac{\delta_I+u_{12}}{\tau},-\frac{1}{\tau},\frac{\sigma}{\tau})$
is holomorphic in our $u_a$ domain.
We can make a similar analysis for
\begin{equation}\label{Gamma-tau-vector}
  \Gamma({\textstyle
  \frac{u_{12}}{\tau},-\frac{1}{\tau},\frac{\sigma}{\tau}})^{-1}
  =\prod_{m,n=0}^\infty
  \frac{1-e^{2\pi i(\frac{u_{12}}{\tau}-\frac{m}{\tau}+\frac{n\sigma}{\tau})}}
  {1-e^{2\pi i(-\frac{u_{12}}{\tau}
  -\frac{m+1}{\tau}+\frac{(n+1)\sigma}{\tau})}}\ .
\end{equation}
This is actually repeating the studies of imaginary parts above, at $\delta=0$.
As for the denominator, one similarly finds its log is holomorphic at no extra
condition. Log of the numerator at given $m,n$ will be holomorphic
if $m+\mathcal{N}(n+x_{12})$ is positive for all
$-1<x_{12}<1$. (We ignore the edges at $x_{12}=\pm 1$, in that no eigenvalues
will sharply assume the edge values in the fine-grained picture.
Anyway, such edge factors will make measure $0$ contributions
in the large $N$ limit.)
This condition is always met except when $m=n=0$.\footnote{One may think
an extra condition $\mathcal{N}<1$ is required for $(m,n)=(1,0)$, but
this is always implied by (\ref{condition-V-tau}).} In other words,
the only factor in the numerator whose log fails to be holomorphic is collected as
(\ref{V-tau-exception}). So, as asserted, the integrand
containing $\tilde{V}_\tau$ of (\ref{potential-1}) can be written as
(\ref{V-tilde-V}), where $V_\tau$ is holomorphic if the conditions
(\ref{condition-V-tau}) are met.

The branch cuts/points of $V_\sigma$ can be studied in a completely
analogous way. After the analysis of the poles/zeros of the functions
$\Gamma(-\frac{\delta+u_{12}+1}{\sigma},-\frac{1}{\sigma},-\frac{\tau}{\sigma})$
and $\Gamma(-\frac{u_{12}+1}{\sigma},-\frac{1}{\sigma},-\frac{\tau}{\sigma})$,
one finds that there are no branch cuts if we assume
\begin{equation}\label{condition-V-sigma}
  \frac{{\rm Im}(-\frac{\tau}{\sigma})}{{\rm Im}(-\frac{1}{\sigma})}
  <\min\left[\frac{1-a_I}{1-b_I},\frac{a_I}{b_I}\right]\ .
\end{equation}
This follows by repeating the analysis of the previous paragraph.

To summarize, the Yang-Mills matrix model in the sector
\begin{equation}\label{sector-1}
  \delta_1+\delta_2+\delta_3=\sigma+\tau-1\ ,\ \
  \delta_I=-a_I+b_I(\sigma+\tau)\ \ \textrm{with}\ \
  a_I,b_I\in(0,1)
\end{equation}
at ${\rm Im}(\frac{\sigma}{\tau})>0$
can be written as
\begin{equation}\label{YM-final}
  Z=\exp\left[-\frac{\pi iN^2\delta_1\delta_2\delta_3}{\sigma\tau}\right]
  \int\prod_{a=1}^N du_a\cdot\frac{1}{N!}
  \prod_{a\neq b}(1-e^{\frac{2\pi iu_{ab}}{\tau}})\cdot
  \exp\left[-\sum_{a\neq b}\left(V_\sigma(u_{ab})+V_\tau(u_{ab})\right)\right]\ ,
\end{equation}
which is basically (\ref{toy-full-haar}) multiplied by a prefactor.
$V_\sigma$, $V_\tau$ satisfy all the required conditions (periods,
holomorphicity) supposing that the chemical potentials satisfy
the conditions (\ref{condition-V-tau}) and (\ref{condition-V-sigma}),
which we write more intrinsically as
\begin{equation}\label{condition-intrinsic}
  {\rm Im}\left(\frac{\sigma-\delta_I}{\tau}\right)<0\ ,\ \
  {\rm Im}\left(\frac{1+\delta_I}{\tau}\right)<0\ ,\ \
  {\rm Im}\left(\frac{1-\tau+\delta_I}{\sigma}\right)<0\ ,\ \
  {\rm Im}\left(\frac{\delta_I}{\sigma}\right)>0\ .
\end{equation}
In the same sector (\ref{sector-1}) at
${\rm Im}(\frac{\sigma}{\tau})<0$ (i.e. ${\rm Im}(\frac{\tau}{\sigma})>0$),
one rewrites the matrix model by using a modular identity
which exchanges the role of $\sigma,\tau$, arriving at a matrix model of
the form (\ref{YM-final}) with $\sigma,\tau$ flipped. The corresponding
$V_\sigma,V_\tau$ again satisfy periodicities and holomorphy when
the $\sigma\leftrightarrow\tau$ flipped version of
(\ref{condition-intrinsic}) is met. So applying the
identity (\ref{haar-identity}) with either $\kappa=\frac{1}{\tau}$ or
$\frac{1}{\sigma}$ to set up the saddle point problem,
one finds the uniform parallelogram
saddle point with the two edges given by $\sigma,\tau$.

The complex conjugate sector
\begin{equation}\label{sector-2}
  \delta_1+\delta_2+\delta_3=\sigma+\tau+1\ ,\ \
  \delta_I=-a_I+b_I(\sigma+\tau)\ \ \textrm{with}\ \
  -a_I,b_I\in(0,1)
\end{equation}
can be studied similarly, by using the second modular transformation
of (\ref{modular}).
If ${\rm Im}(\frac{\sigma}{\tau})>0$, one obtains
\begin{equation}\label{YM-final-2}
  Z=\exp\left[-\frac{\pi iN^2\delta_1\delta_2\delta_3}{\sigma\tau}\right]
  \int\prod_{a=1}^N du_a\cdot\frac{1}{N!}
  \prod_{a\neq b}(1-e^{\frac{2\pi iu_{ab}}{\sigma}})\cdot
  \exp\left[-\sum_{a\neq b}\left(V_\sigma(u_{ab})+V_\tau(u_{ab})\right)\right]\ ,
\end{equation}
where the potentials $V_\sigma,V_\tau$ are periodic and holomorphic if
\begin{equation}
  \frac{{\rm Im}(\frac{\sigma}{\tau})}{{\rm Im}(-\frac{1}{\tau})}
  <\min\left[\frac{1-|a_I|}{1-b_I},\frac{|a_I|}{b_I}\right]\ ,\ \
  \frac{{\rm Im}(\frac{-\tau}{\sigma})}{{\rm Im}(-\frac{1}{\sigma})}
  <\min\left[\frac{|a_I|}{1-b_I},\frac{1-|a_I|}{b_I}\right]
\end{equation}
or more intrinsically
\begin{equation}\label{condition-intrinsic-2}
  {\rm Im}\left(\frac{\delta_I-\tau}{\sigma}\right)<0\ ,\ \
  {\rm Im}\left(\frac{1-\delta_I}{\sigma}\right)<0\ ,\ \
  {\rm Im}\left(\frac{1+\sigma-\delta_I}{\tau}\right)<0\ ,\ \
  {\rm Im}\left(\frac{\delta_I}{\tau}\right)<0\ .
\end{equation}
Similar matrix model can be derived at ${\rm Im}(\frac{\sigma}{\tau})<0$,
with the roles of $\sigma,\tau$ flipped. So again we
find the saddle given by the uniform parallelogram distribution
with edges $\sigma,\tau$.

We also consider the large $N$ free energy at our
parallelogram saddle point
\begin{equation}\label{free-para}
  \log Z= -\frac{\pi i N^2\delta_1\delta_2\delta_3}{\sigma\tau}
  -N^2\int_{-\frac{1}{2}}^{\frac{1}{2}}
  dx_1 dy_1 dx_2 dy_2\left[V_\sigma(\sigma x_{12}+\tau y_{12})+
  V_\tau(\sigma x_{12}+\tau y_{12})\right]
\end{equation}
in all cases summarized above,
where one of $V_\sigma,V_\tau$ includes the
half-Haar-like measure, which we called $V_\sigma^{(\pm)}$ or $V_\tau^{(\pm)}$
in (\ref{V-pm}).
We first evaluate
\begin{equation}
  \int_{-\frac{1}{2}}^{\frac{1}{2}}
  dx_1 dy_1\left[V_\sigma(u_{12})+V_\tau(u_{12})\right]
\end{equation}
at fixed $u_2$. As for the integral of $V_\sigma$, we compute the integrals
of the form
\begin{equation}
  -\int_{-\frac{1}{2}}^{\frac{1}{2}}
  d x_1 \log\left(1-f(u_2,y_1) e^{\pm 2\pi i x_{12}}\right)\ ,
\end{equation}
where $|f|<1$ in the whole integration domain. This can be computed by
Taylor-expanding the integrand, since the integral domain is within
the radius of convergence:
\begin{equation}
  \sum_{n=1}^\infty\frac{f(u_2,y_1)^n}{n}\int_0^1 dx_1 e^{\pm 2\pi i x_{12}}=0\ .
\end{equation}
Similarly, the log terms contained in $V_\tau$
trivially integrates to zero for the same reason,
by Taylor-expanding and integrating over $y_1$ first.
So the contribution from the second term of (\ref{free-para})
vanishes. The large $N$ free energy of our parallelogram saddle point
is thus given by
\begin{equation}\label{free-final}
  \log Z=-\frac{\pi i N^2\delta_1\delta_2\delta_3}{\sigma\tau}
  =\frac{N^2\Delta_1\Delta_2\Delta_3}{2\omega_1\omega_2}\ ,
\end{equation}
in both sectors with $\sum_I\delta_I=\sigma+\tau\mp 1$ (or
$\sum_I\Delta_I=\omega_1+\omega_2\pm 2\pi i$).
This completely accounts for the entropy function of the BPS black holes
in $AdS_5\times S^5$ discovered in \cite{Hosseini:2017mds}.
The two sectors of (\ref{delta-classify}) with upper/lower signs provide saddles in
the mutually complex conjugate regions. This paired structure
plays important roles to realize the macroscopic index with oscillating signs
upon Legendre transformation to microcanonical ensemble \cite{Agarwal:2020zwm}.

The conditions (\ref{condition-intrinsic}) or
(\ref{condition-intrinsic-2}) are nontrivial if $\sigma,\tau$ are not
collinear. Let us first show that
these conditions are trivially met in the collinear limit,
$\frac{\sigma}{\tau}=$ real. For simplicity, here we just discuss the first case
(\ref{condition-intrinsic}). At $\frac{\sigma}{\tau}=$ real,
these conditions reduce to
\begin{equation}
  {\rm Im}\left(\frac{\delta_I}{\tau}\right)>0\ ,\ \
  {\rm Im}\left(\frac{1+\delta_I}{\tau}\right)<0\ ,
\end{equation}
which together with ${\rm Im}(\tau)>0$ demand
$0<a_I<1$. This is the condition already met in the upper sector of
(\ref{delta-classify}), basically set by the periodic shifts of $\delta_I$'s.
Therefore, the condition (\ref{condition-intrinsic}) is always met
in the collinear limit.

If $\sigma,\tau$ are non-collinear, (\ref{condition-intrinsic})
or (\ref{condition-intrinsic-2}) nontrivially constrain the parameters
$\delta_I,\sigma,\tau$ for our basic parallelogram to solve the saddle point
equation. We shall analyze in appendix A what kind of constraints are imposed by
these conditions. These conditions are very reminiscent of the stability conditions
of the Euclidean black hole solutions against D3-brane instantons wrapping
$S^3\subset S^5$ and $S^1\subset S^3\subset AdS_5$ \cite{Aharony:2021zkr}.
Their contributions to the partition function is given by
$Z\leftarrow e^{iS_{\textrm{D3}}}$ with
$S_{\textrm{D3}}=\pm 2\pi N\frac{\delta_I}{\sigma}$ or
$\pm 2\pi N\frac{\delta_I}{\tau}$ in the two cases of (\ref{delta-classify}).
Our conditions imply their stability conditions
${\rm Im}(S_{\textrm{D3}})>0$, which makes our results
consistent with the gravity analysis. Our conditions
are stronger than their stability conditions, leading us to
conjecture that there are more stability constraints from
other instantons not discussed in \cite{Aharony:2021zkr}.
Since our conditions (\ref{condition-intrinsic}), (\ref{condition-intrinsic-2})
come from the requirement that no branch points of the potential are
included in the parallelogram, it would be interesting to establish
a direct connection between the forces generated by the branch points
and the gravitational instability. It would be interesting to
study the parameter regime outside the conditions (\ref{condition-intrinsic})
or (\ref{condition-intrinsic-2}), especially the physics of the
corresponding black holes.
For instance, the thermodynamic instability of small spinning black holes
have been already discussed in \cite{Choi:2021lbk}.

We can also construct generalized parallelogram saddles with the edge vectors
given by $(\sigma+r,\tau+s)$, where $r,s\in\mathbb{Z}$. To discuss these generalized
solutions, we set the
convention for $\delta_I$'s so that one of the two conditions (\ref{delta-classify})
is met with $\sigma,\tau$ replaced by $\sigma+r,\tau+s$. Then repeating the
calculus of this section,
one finds that the parallelogram ansatz with edges given by
$\sigma+r,\tau+s$ solves the saddle point equation
if (\ref{condition-intrinsic}) or (\ref{condition-intrinsic-2}) is met after
replacing $\sigma,\tau$ by $\sigma+r,\tau+s$. In appendix A, we show that
(\ref{condition-intrinsic}) or (\ref{condition-intrinsic-2}) is always met by
many choices of $r,s$, at least for typical choice of $\sigma,\tau$ which
satisfies $\frac{\sigma}{\tau}\neq$ real and
$\frac{{\rm Im}(\sigma)}{{\rm Im}(\tau)}\neq $ rational.
Following \cite{Aharony:2021zkr}, we interpret our $r,s$ as labelling
multiple Euclidean solutions which map to the same Lorentzian solution
once we compactify the temporal circle.

\section{Alternative derivation in the collinear limit}

Our analysis so far demands that $\sigma$, $\tau$ are not collinear,
at least at the intermediate steps. Otherwise the
modular parameters $\frac{\sigma}{\tau}$, $-\frac{\tau}{\sigma}$
in the identity (\ref{modular}) become real and make individual
$\Gamma$ functions ill defined. However, the full
integrand is smooth at real $\frac{\sigma}{\tau}$.
So even if we are interested
in the case with collinear $\sigma,\tau$, we may slightly deform them
with small ${\rm Im}(\frac{\sigma}{\tau})$
and remove this regulator after the calculations. This way, one obtains
linear cut distributions for collinear $\sigma,\tau$.
As shown in Fig. \ref{parallel-cartoon}, the
eigenvalue distribution along the linear cut is no longer uniform.
We can parametrize the eigenvalues as $u(x)=\frac{\sigma+\tau}{2} x$ with
$x\in(-1,1)$ in the collinear limit.
Defining real $R\equiv\frac{\sigma}{\tau}$,
the linear eigenvalue density $\rho(x)$ is given by
\begin{equation}\label{trapezoid}
  \rho(x)=\frac{1}{1-(\frac{1-R}{1+R})^2}
  \left(1-\frac{1}{2}\left| x+\frac{1-R}{1+R}\right|
  -\frac{1}{2}\left| x-\frac{1-R}{1+R}\right|\right)\ .
\end{equation}
$\rho(x)$ satisfies $\int_{-1}^1dx\rho(x)=1$, and
the eigenvalue sum is replaced by $\sum_a\rightarrow N\int_{-1}^1dx\rho(x)$.
As shown in Fig. \ref{parallel-cartoon}(b), $\rho(x)$ has a
trapezoid shape. We hope this shape is intuitively visible
from the degeneration shown in Fig. \ref{parallel-cartoon}(a). This $\rho(x)$
has been already derived in the small black hole limit at
real $\frac{\sigma}{\tau}$ \cite{Choi:2021lbk}, using the standard large $N$
matrix model techniques. In the sector given by the upper signs of
(\ref{delta-classify}), the small black hole limit is given by
\begin{equation}
  \sigma\rightarrow \frac{1}{2}+\frac{i\gamma}{2\pi}\ ,\ \
  \tau\rightarrow \frac{1}{2}-\frac{i\gamma}{2\pi}
\end{equation}
in the notation of \cite{Choi:2021lbk}.
(The imaginary part of $\sigma+\tau$ approaches zero because small black hole
limit is a kind of high temperature limit.)
When $\sigma=\tau$, or $R=1$ ($J_1=J_2$),
(\ref{trapezoid}) becomes triangular. The triangular distribution
was found in the small black hole limit \cite{Choi:2021lbk},
and also from the subleading correction to the Cardy limit for large
black holes \cite{GonzalezLezcano:2020yeb}.

Here we emphasize that the linear distributions at collinear $\sigma,\tau$ were
found in special cases without changing the saddle point problem using the identity (\ref{haar-identity}) \cite{Choi:2021lbk,GonzalezLezcano:2020yeb}.
In fact, one can make an alternative derivation of our linear distributions
at collinear $\sigma,\tau$, without changing the integrand
using (\ref{haar-identity}) and also without using the modular identity
(\ref{modular}).

For simplicity, we only consider the upper case of (\ref{delta-classify})
in this section. The 2-body eigenvalue potential (including the Haar measure)
is given by
\begin{equation}\label{potential}
  -V(u)=\frac{1}{2}\sum_{I=1}^3\log\Gamma(\delta_I+u;\sigma,\tau)
  -\frac{1}{2}\log\Gamma(u;\sigma,\tau)+
  \left(u\rightarrow -u\right)\ .
\end{equation}
As before, there is an issue of how we define $\log$ functions
concerning the choice of branch sheets on the $u$ space.
In the full discrete setup, they only affect $2\pi i\mathbb{Z}$
constants of $\log Z$ and is thus irrelevant. In the continuum limit,
the most natural and useful setup is to define the log functions to
be smooth on the eigenvalue cut of our ansatz, so that their
derivatives (force) are well defined everywhere.
As we explain in a moment, with a careful definition of our ansatz,
this will be possible everywhere except when two eigenvalues
approach each other, where one finds the usual repulsive
singularity. We know how to continue the log functions across such
repulsive singularities, by making a principal-valued definition of
log functions. Therefore, we shall be able to define the log functions
based on continuity on the eigenvalue cut.

Now we discuss our ansatz for collinear $\sigma,\tau$ in more detail,
in the original saddle point problem without using
the identity (\ref{haar-identity}).
First of all, it is still convenient to parametrize the
eigenvalues $u_a$'s using
two parameters, $u(x,y)=x\sigma+y\tau$ with $x,y\in(-\frac{1}{2},\frac{1}{2})$
and $\rho(x,y)=1$. This is no longer a regular parallelogram,
but we can label the eigenvalues this way.
A point on the physical eigenvalue cut in the $u$ space
is mapped to a segment in the square region $x,y\in(-\frac{1}{2},\frac{1}{2})$.
$\log Z$ is
given by a double sum $\sum_{a,b=1}^N$, which in the large $N$
continuum limit is replaced by the double integral
\begin{equation}
  \log Z=-N^2\int dx_1dy_1 \int dx_2dy_2 V(u_1-u_2)\ .
\end{equation}
One finds $u=x\sigma+y\tau=\tau\left(Rx+y\right)$
with a real $R$. Regarding $x$ and $\tilde{y}\equiv y+Rx$ as
the integral variables, one can first trivially integrate over $x$
at fixed $\tilde{y}$. The allowed ranges of $x$ at given $\tilde{y}$
determine the linear eigenvalue density $\rho(\tilde{y})$. For instance,
if $R<1$, they are given by
\begin{eqnarray}
  0<\tilde{y}<R&:&
  {\textstyle 0<x<\frac{\tilde{y}}{R}\ \rightarrow
  \ \rho(\tilde{y})=\frac{\tilde{y}}{R}}\\
  R<\tilde{y}<1&:&0<x<1\ \rightarrow\ \rho(\tilde{y})=1\nonumber\\
  1<\tilde{y}<1+R&:&{\textstyle \frac{\tilde{y}-1}{R}<x<1
  \ \rightarrow\ \rho(\tilde{y})=1-\frac{\tilde{y}-1}{R}}\nonumber
\end{eqnarray}
which is (\ref{trapezoid}) upon reparametrizing $\tilde{y}$.
Here we emphasize that our collinear ansatz with the original
potential (without using (\ref{haar-identity})) will demand a small refinement
to make it a saddle point. Namely, we take the precise ansatz to be
\begin{equation}\label{ansatz-epsilon}
  u(x,y)=e^{-i\epsilon}(x\sigma+y\tau)=\tau e^{-i\epsilon}(Rx+y)
  \ ,\ \ \rho(x,y)=1\ \ \ ({\textstyle -\frac{1}{2}<x,y<\frac{1}{2}})
\end{equation}
with `infinitesimal' $\epsilon>0$. During most of the calculus, we can simply
turn off $\epsilon=0$. Below we shall keep small $\epsilon>0$ only when
necessary.

Employing the ansatz $u_{1,2}=\sigma x_{1,2}+\tau y_{1,2}+\mathcal{O}(\epsilon)$,
it will be useful to rewrite the $u_1$
integral of $\log \Gamma(z\pm u_{12})$ as
\begin{eqnarray}\label{log-gamma}
  &&\int_{-\frac{1}{2}}^{\frac{1}{2}}dx_1dy_1
  \log\Gamma(z\pm u_{12},\sigma,\tau)\\
  &&=\int_{-\frac{1}{2}}^{\frac{1}{2}}dx_1 dy_1
  \sum_{m,n=0}^\infty\left[-\log(1-e^{2\pi i(\pm u_1+m\sigma+n\tau\mp u_2+z)})
  +\log(1-e^{2\pi i(\mp u_1+m\sigma+n\tau+\sigma+\tau-z\pm u_2)})\right]
  \nonumber\\
  &&=\int_{-\frac{1}{2}}^{\infty} dxdy
  \left[-\log(1-e^{2\pi i(u\mp u_2+z)})
  +\log(1-e^{2\pi i(u+\sigma+\tau-z\pm u_2)})\right]\ .\nonumber
\end{eqnarray}
On the last line, we defined $u=\sigma x +\tau y$ as
$u\equiv\pm u_1+m\sigma+n\tau$,
with $x\in(-\frac{1}{2}+m,\frac{1}{2}+m)$,
$y\in(-\frac{1}{2}+n,\frac{1}{2}+n)$, and patched the infinitely many
integrals labeled by $m,n$ into a single integral over
$x,y\in(-\frac{1}{2},\infty)$ (at $\epsilon\rightarrow 0$).
This integral is a useful object for a couple of reasons.
Firstly, if we further integrate once more with $x_2,y_2$, it will give
the contribution of each supermultiplet to the free energy $\log Z$.
Also, taking $u_2$ derivative, one obtains
a force acting on the eigenvalue located at $u_2$.
The natural and useful convention
for the log functions on the last line is to define them
as a continuous function of $x,y\in (-\frac{1}{2},\infty)$,
except for the principal-valued singularity when $z=0$ and $u\mp u_2=0$.
We shall call this the Haar measure singularity below.
We now show that such continuous definitions of log are possible.

At large $x,y$, we naturally stay on a branch of log function
which yields $\log 1=0$ at $x,y\rightarrow\infty$, and
attempt to define the log function as a continuous function in the whole $x,y$ domain,
if necessary by moving on to different branch sheets.
Although we employ a 2 dimensional parametrization, the eigenvalues
are on a linear cut (labeled by $Rx+y$) so that there will be no
ambiguities in 1 dimension to make such a continuous extension except at
the Haar measure singularity. We shall establish the details of
this continuation below.

Firstly, when $z=\delta_I$,
$u\mp u_2+\delta_I$ and
$u+\sigma+\tau-\delta_I\pm u_2$
have positive imaginary part for large $x,y$, in which case
the two log functions on the last line of
(\ref{log-gamma}) are defined on the standard branch with $\log 1=0$.
As long as their imaginary parts are positive, the log function can
be defined by the standard Taylor expansion
$\log(1-x)=-\sum_{n=1}^\infty\frac{x^n}{n}$.
We study when this definition has to be modified
by analytic continuation, by the imaginary parts changing sign.
For the first log on the last line of (\ref{log-gamma}), the sign
changes at the following line on the $x,y$ space:
\begin{equation}\label{matter-first-log}
  0={\rm Im}(u\mp u_2+\delta_I)
  ={\rm Im}(\tau)\left[R(x+b_I\mp x_{2})+
  (y+b_I\mp y_{2})\right]\ .
\end{equation}
(When $z=\delta_I$, the refined definition (\ref{ansatz-epsilon})
plays no role.) On this line, its real part
\begin{equation}
  {\rm Re}(u\mp u_2+\delta_I)=-a_I+{\rm Re}(\tau)
  \left[R(x+b_I\mp x_{2})+(y+b_I\mp y_{2})\right]=-a_I
\end{equation}
is in the range $-1<-a_I<0$. In particular,
the branch point is never crossed when analytic continuation is needed
to define this log. So when $z=\delta_I$,
the first log function of (\ref{log-gamma}) in the
region ${\rm Im}(u\mp u_2+\delta_I)<0$ is continuously extended using the
formula
\begin{equation}\label{log-continue}
  \log(1-e^{2\pi iz})=\log(1-e^{-2\pi iz})+2\pi iz+\pi i\ \ \
  \textrm{ if }\  {\rm Im}(z)<0\textrm{ and } \ {\rm Re}(z)\in (-1,0)\ .
\end{equation}
The log on the right hand side is defined by Taylor expansion.
Similarly, as for the second log of (\ref{log-gamma}),
we regard $-\delta_I$ as $-\delta_I-1$ in the infinity branch, which
is just a phase convention for the fugacity.
Then one finds ${\rm Im}(u+\sigma+\tau\pm u_2-\delta_I-1)=0$ at
\begin{equation}
  0={\rm Im}(\tau)\left[R(x-b_I\pm x_{2}+1)+
  (y-b_I\pm y_{2}+1)\right]\ .
\end{equation}
Then its real part is given by
\begin{equation}
  a_I-1+{\rm Re}(\tau)
  \left[R(x-b_I\pm x_{2}+1)+(y-b_I\pm y_{2}+1)\right]=a_I-1
\end{equation}
which is in the range $-1<a_I-1<0$. So we can again use
(\ref{log-continue}) to define this log by analytic continuation,
after replacing $-\delta_I$ by $-\delta_I-1$.

When $z=0$, only the first log of (\ref{log-gamma}) can hit the branch point,
while the second log can always be defined by Taylor expansion. For the first log,
the branch with $\log 1=0$ chosen at infinity extends smoothly to the region
${\rm Im}(u\mp u_2)>0$, whose boundary is the line
${\rm Im}(u\mp u_2)=0$. Had we defined our ansatz as
$u(x,y)=\sigma x+\tau y$ rather than (\ref{ansatz-epsilon}),
the real and imaginary parts
of $u\mp u_2$ become zero at the same point,
\begin{equation}
  Rx+y=\pm (Rx_2+y_2)\ ,
\end{equation}
meaning that the branch point is on the eigenvalue cut.
Here our refined definition (\ref{ansatz-epsilon}) has a finite
effect even at $\epsilon\rightarrow 0^+$.
Since $u\equiv \pm u_{12} + m\sigma+n\tau$ with
$m,n\geq 0$, one finds
\begin{equation}
  u\mp u_2=\tau \left[\pm e^{-i\epsilon}(Rx_{12}+y_{12})+mR+n\right]\ .
\end{equation}
Namely, due to $\epsilon>0$ in our ansatz, the real and
imaginary parts of $u\mp u_2$ do not simultaneously vanish
unless $Rx_{12}+y_{12}=0$ and $m=n=0$. The last point is the Haar measure
singularity. Our $\epsilon$ deforms other branch point singularities
slightly away form the eigenvalue cut.

\begin{figure}[!t]
\centering
\includegraphics[width=0.5\textwidth]{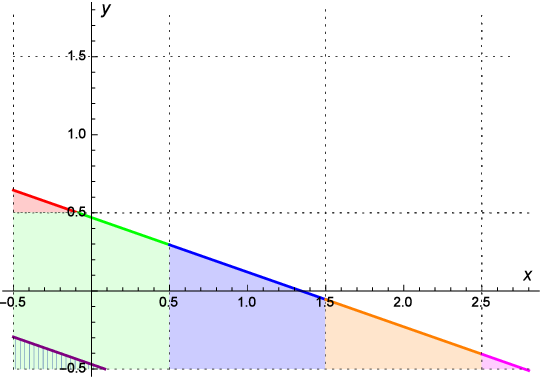}
\caption{Singular points of the potential for the naive ansatz at $\epsilon=0$,
from $u=-u_2$ (purple line) and $u=u_2$ (red/green/blue/orange/magenta).
Each square region bounded by dotted lines is a fundamental region of
$(x_1,y_1)$, whose integrand is the log potential at certain $(m,n)$.
Each colored line segment represents a point on the eigenvalue cut.
Green/purple lines are the Haar measure singularities,
which remain after the ansatz deformation by $\epsilon$. Other singularities
(red, blue, orange, magenta) are lifted after the deformation.
($R\equiv\frac{\sigma}{\tau}=.35$, $Rx_2+y_2=.47$)}
\label{singularity}
\end{figure}

Let us explain this in detail.
On the quadrant defined by $x,y\in (-\frac{1}{2},\infty)$, the line
after which analytic continuation is needed is
\begin{eqnarray}\label{vector-imaginary}
  0&=&{\rm Im}(u\mp u_2)={\rm Im}(\tau)
  \left[\pm \cos\epsilon(Rx_{12}+y_{12})+mR+n\right]
  \mp{\rm Re}(\tau)\sin\epsilon(Rx_{12}+y_{12})\nonumber\\
  &\approx&{\rm Im}(\tau)\left[\pm(Rx_{12}+y_{12})+mR+n\right]
  \mp\epsilon{\rm Re}(\tau)(R_{12}+y_{12})
\end{eqnarray}
up to linear order in $\epsilon$.
The leading $\mathcal{O}(\epsilon^0)$ shape of these lines
$Rx_1+y_1=Rx_2+y_2\mp(Rm+n)$, or $Rx+y=\pm(Rx_2+y_2)$, are shown
on the $(x,y)$ space in Fig. \ref{singularity}.
On this line, the real part of
$u\mp u_2$ is given by
\begin{eqnarray}\label{vector-real}
  {\rm Re}(u\mp u_2)&=&{\rm Re}(\tau)
  \left[\pm \cos\epsilon(Rx_{12}+y_{12})+mR+n\right]
  \pm{\rm Im}(\tau)\sin\epsilon(Rx_{12}+y_{12})\nonumber\\
  &\approx&\pm\epsilon\left({\textstyle \frac{{\rm Re}(\tau)^2}{{\rm Im}(\tau)}}
  +{\rm Im}(\tau)\right)(R x_{12}+y_{12})
\end{eqnarray}
again up to $\mathcal{O}(\epsilon^1)$, where we inserted the condition
${\rm Im}(u\mp u_2)=0$. Therefore, unless $Rx_{12}+y_{12}=0$
(i.e. $u_1=u_2$), a small real
part of $u\mp u_2$ is generated by $\epsilon$
on the line (\ref{vector-imaginary}), represented by the
red/blue/orange/magenta parts of the line in Fig. \ref{singularity}.

We explain the situation of Fig. \ref{singularity} in
more detail. For illustration, we chose $R=0.35$ and $Rx_2+y_2=.47>0$.
The two lines with green/purple colors in the region
$-\frac{1}{2}<x,y<\frac{1}{2}$ represent the
Haar measure singularities, for which ${\rm Re}(u\mp u_2)=0$ exactly.
These are the only branch points on the eigenvalue cut at nonzero $\epsilon$,
for which we shall review in a moment how the functions are extended across
the singularity (in a standard manner).
The regions requiring continuations across these lines are
also shown as shades with the same colors. Other line segments in
red/blue/orange/magenta colors for ${\rm Im}(u-u_2)=0$, are
for the line $Rx_1+y_1\approx (Rx_2+y_2)-(Rm+n)$
with $(m,n)=(0,1), (1,0),(2,0),(3,0)$ respectively at $\mathcal{O}(\epsilon^0)$.
For these, note that $Rx_{12}+y_{12}=-(Rm+n)<0$. So from (\ref{vector-real}),
all parts of the line (\ref{vector-imaginary}) except for the Haar measure
singularities (green/purple) have small negative
${\rm Re}(u-u_2)\sim\mathcal{O}(\epsilon^1)<0$.
Since $-1<{\rm Re}(u-u_2)<0$, one can apply the analytic continuation
(\ref{log-continue}) across these lines to the regions shaded with the
same colors.

So except for the Haar measure contributions, which is
the integral (\ref{log-gamma}) in the domain $x,y\in(-\frac{1}{2},\frac{1}{2})$
(the square region including green/purple segments in Fig. \ref{singularity}),
the log functions in (\ref{log-gamma}) are defined either by the Taylor expansion
or the continuation (\ref{log-continue}). The Haar measure contribution
\begin{equation}\label{integral-haar}
  \int_{-\frac{1}{2}}^{\frac{1}{2}} dxdy
  \left[\log(1-e^{2\pi i(\sigma(x-x_2)+\tau(y-y_2))})
  +\log(1-e^{2\pi i(\sigma(x+x_2)+\tau(y+y_2))})\right]
\end{equation}
has to be treated as the principal-valued integral.
One way of doing this calculus is to eliminate the $\varepsilon$
neighborhood of the singularity, and send $\varepsilon\rightarrow 0^+$ after
the calculation. Another equivalent way is to
average over the $\pm i\varepsilon$ deformations of the integration contour.
This amounts to averaging over the integral done with the analytic continuation
(\ref{log-continue}) and another integral with alternative continuation
\begin{equation}\label{log-continue-2}
  \log(1-e^{2\pi iz})=
  \log(1-e^{-2\pi iz})+2\pi iz-\pi i\ \ \
  \textrm{ if }\  {\rm Im}(z)<0\textrm{ and } \ {\rm Re}(z)\in(0,1)\ .
\end{equation}
The two calculations differ by whether one uses $\pm \pi i$ on the
last terms of (\ref{log-continue}) and (\ref{log-continue-2}),
in the region shaded with green/purple colors in Fig. \ref{singularity}.
So they are just different by integrating
constants over these regions. If one averages over the two, the integrations
of $\pm \pi i$ cancels. Therefore, (\ref{integral-haar}) computed using
(\ref{log-continue}) and (\ref{log-continue-2}) prescriptions are related to
the principal-valued integration of (\ref{integral-haar}) by having
the following additional constants, respectively:
\begin{equation}
  \pm \pi i\left[\textrm{area}(\textrm{green shaded region})+
  \textrm{area}(\textrm{purple shaded region})\right]=\pm \pi i\ .
\end{equation}
At the last step we used the fact that the sum of the areas of the two regions
is equal to the area of the square $-\frac{1}{2}<x,y<\frac{1}{2}$, which is $1$.
Therefore, even for the Haar measure integral
we may employ a unified prescription to analytically continue with
(\ref{log-continue}), and add a trivial constant $-\pi i$ to get the
principal-valued integral for the Haar measure potential.
In fact for most purposes, we can ignore this constant $-\pi i$.
For the force calculation, this constant factor does not matter. Also,
for the free energy calculation, this constant will provide an
extra imaginary constant $-\frac{\pi i(N^2-N)}{2}\sim-\frac{\pi iN^2}{2}$
to the free energy $\log Z$. This factor provides an overall sign factor
(or phase factor) for $Z\leftarrow (-1)^{\frac{N^2-N}{2}}$,
but otherwise does not affect the large $N$ thermodynamics. So
we ignore this constant term from now and proceed by universally employing
the analytic continuation (\ref{log-continue}) for (\ref{log-gamma}).

We have set all the rules of calculus in the collinear case.
In a moment we will show that the force vanishes, and then
compute the free energy. Before the calculations,
we pause to interpret what small $\epsilon>0$ may mean.
Certainly $\epsilon$ is part of our ansatz. In our leading large $N$ calculus,
only the sign of $\epsilon$ will matter. Our ansatz (\ref{ansatz-epsilon})
is a saddle point for $\epsilon>0$, but not for $\epsilon<0$.
Regarding it literally as an infinitesimal parameter appears to be
unrealistic, since it measures the distance of a potential singularity from
the eigenvalue configuration and it cannot happen in the discrete calculus
that the saddle point is infinitesimally away from a point where the force diverges.
So we interpret infinitesimal $\epsilon$ as emerging from the large $N$
continuum limit. For instance, if one can make a subleading calculus in $\frac{1}{N}$,
it may be related to $N$ by $\epsilon\sim\frac{1}{N^\alpha}$ with a positive number
$\alpha$. (We expect $\alpha<1$, for $\epsilon$ to be larger than the minimal
eigenvalue separation $\sim \frac{1}{N}$.)
Of course if one can actually do a subleading calculus, the saddle will be
more complicated than (\ref{ansatz-epsilon}). Our (\ref{ansatz-epsilon})
merely prescribes how the branch point is avoided at large $N$.
Here we note that a similar $\epsilon$
deformation was needed to get the saddle point of this index in the Cardy limit
\cite{GonzalezLezcano:2020yeb}. We expect their $\epsilon$ should be interpreted
similarly, as a small number related to the large charges.

More physically, the singularities
which are $\epsilon$-distance away from our ansatz come from the
gaugino operators dressed by derivatives. In the notation of
\cite{Grant:2008sk}, the gaugino `letter'
$(\partial_{+\dot{+}})^{p}(\partial_{+\dot{-}})^{q}\bar{\lambda}_{\dot\pm}$
in the $a$'th row and $b$'th column of the $N\times N$ matrix is
weighted by the following effective fugacity factor in the matrix integral:
\begin{equation}\label{gaugino-fugacity}
  e^{2\pi iu_{ab}}\cdot
  e^{2\pi i\sigma(\frac{1}{2}\pm\frac{1}{2}+p)}
  e^{2\pi i\tau(\frac{1}{2}\mp\frac{1}{2}+q)}\ .
\end{equation}
This is a product of the `color fugacity' factor
$e^{2\pi iu_{ab}}$ which is not a physical fugacity, and the rest
which is the physical fugacity.
Since $u_a$ in our ansatz is on the straight line interval
$(-\frac{\sigma+\tau}{2},\frac{\sigma+\tau}{2})$ in the $\epsilon=0$ limit,
$u_{ab}$ is on the interval $(-(\sigma+\tau),\sigma+\tau)$.
Therefore, although the physical fugacity has its absolute value smaller
than $1$, the factor $e^{2\pi iu_{ab}}$ may be larger than $1$
and make (\ref{gaugino-fugacity}) close to $1$ for certain $u_a,u_b$.
If this happens, the potential $V(u_{ab})$ will almost diverge.
The integrand of the vector multiplet can be simplified as
\begin{equation}
  \prod_{a\neq b}\Gamma(u_{ab},\sigma,\tau)^{-1}=\exp\left[
  -\sum_{a\neq b}\sum_{p=1}^\infty\frac{1}{p}e^{2\pi ipu_{ab}}
  \left(1+{\textstyle \sum_{m=1}^\infty} e^{2\pi ipm\sigma}+
  {\textstyle \sum_{n=1}^\infty}e^{2\pi ipn\tau}\right)\right]\ .
\end{equation}
The term $1$ in the exponent comes from the Haar measure.
The terms $e^{2\pi ipm\sigma}$ and $e^{2\pi ipn\tau}$ come
from the letters $(\partial_{+\dot{+}})^{m-1}\bar\lambda_{\dot{+}}$
and $(\partial_{+\dot{-}})^{n-1}\bar\lambda_{\dot{-}}$, respectively.
They are responsible for the divergences which are $\epsilon$-distance
away from the red/blue/orange/magenta lines of Fig. \ref{singularity},
labeled by either $(m,0)$ or $(0,n)$ with $m,n\neq 0$. So our saddle
point configuration is very close the point where these charged operators become
massless. The fact that these fermionic operators are very light
(with mass at order $\epsilon$) at our saddle point may provide important
clues on the microstates of the dual black holes or
further generalizations to hairy black holes.
Interestingly, a Fermi surface model for these black holes has been
proposed in \cite{Berkooz:2006wc,Berkooz:2008gc}, precisely based on using
the gaugino letters discussed above. Although the simplest operators of
\cite{Berkooz:2006wc} acquire nonzero anomalous dimensions
above the BPS bound \cite{Berkooz:2008gc}, minor corrections to their ansatz may
be relevant for better understanding the microstates of the BPS black holes.
We hope our findings to provide helpful insights.

We now compute the force and free energy.
To compute both quantities, we first study
\begin{eqnarray}\label{potential-function}
  -N^2\int_{-\frac{1}{2}}^{\frac{1}{2}} dx_1 dy_1 V(u_{12})\!&\!=\!&\!
  \frac{N^2}{2}\sum_{\pm}\int_{-\frac{1}{2}}^\infty dx dy
  \left[\frac{}{}\!\!\log(1-e^{2\pi i(u\mp u_2)})
  -\log(1-e^{2\pi i(u\pm u_2+\sigma+\tau)})\right.
  \nonumber\\
  &&\hspace{-1cm}\left.\frac{}{}\!\!
  -{\textstyle \sum_{I=1}^3}\left(\log(1-e^{2\pi i(u\mp u_2+\delta_I)})
  -\log(1-e^{2\pi i(u\pm u_2+\sigma+\tau-\delta_I-1)})\right)\right]
\end{eqnarray}
where we inserted (\ref{potential}) and (\ref{log-gamma}). Whenever
the analytic continuation has to be made for the log functions, one uses
(\ref{log-continue}) with the exponents as specified in the formula
(including the $-1$ term on the last term). The continuation formula
(\ref{log-continue}) is a special case of the identities for the
polylog functions ${\rm Li}_s(e^{2\pi iz})$. This function is
defined by Taylor expansion
\begin{equation}
  {\rm Li}_s(e^{2\pi iz})=\sum_{n=1}^\infty\frac{e^{2\pi inz}}{n^s}
\end{equation}
when ${\rm Im}(z)>0$, and by the analytic continuation
\begin{equation}\label{polylog-rewrite}
  {\rm Li}_s(e^{2\pi iz})
  =-(-1)^s{\rm Li}_s(e^{-2\pi iz})-\frac{(2\pi i)^s}{s!}B_s(z+1)\ ,
\end{equation}
when ${\rm Im}(z)<0$ and $-1<{\rm Re}(z)<0$.
$B_s(z)$ are the Bernoulli polynomials.
(\ref{log-continue}) is a special case of (\ref{polylog-rewrite}) at
$s=1$, with ${\rm Li}_1(e^{2\pi iz})=-\log(1-e^{2\pi iz})$ and
$B_1(z)=z-\frac{1}{2}$.
These formula are relevant for computing (\ref{potential-function})
since $\frac{\partial}{\partial z}{\rm Li}_s(e^{2\pi iz})
=2\pi i{\rm Li}_{s-1}(e^{2\pi iz})$. After integrating twice with $x$ and $y$,
one obtains
\begin{eqnarray}
  -N^2\int_{-\frac{1}{2}}^{\frac{1}{2}}dx_1dy_1V(u_{12})&=&
  -\frac{N^2}{8\pi^2\sigma\tau}\sum_\pm\left[
  -{\rm Li}_3\left(e^{2\pi i(\mp u_2-\frac{\sigma+\tau}{2})}\right)
  +{\rm Li}_3\left(e^{2\pi i(\frac{\sigma+\tau}{2}\pm u_2)}\right)
  \right.\\
  &&\left.+{\textstyle \sum_{I=1}^3}
  \left({\rm Li}_3\left(e^{2\pi i(\delta_I\mp u_2-\frac{\sigma+\tau}{2})}\right)
  -{\rm Li}_3\left(e^{2\pi i(\frac{\sigma+\tau}{2}\pm u_2-\delta_I-1)}\right)
  \right)\right]\ .\nonumber
\end{eqnarray}
Any ${\rm Li}_3$ functions are defined by the right hand side of
(\ref{polylog-rewrite}) at $s=3$ when they need analytic continuations, since
they are obtained by integrating (\ref{log-continue}).
On the first line, the second term is always defined by Taylor expansion
while the first term needs to be defined by the right hand side of
(\ref{polylog-rewrite}). On the second line, it is always that one of
the two terms is defined by Taylor expansion while the other term is
defined by the right hand side of (\ref{polylog-rewrite}). In both cases,
one universally obtains the following expression:
\begin{eqnarray}\label{potential-constant}
  \hspace*{-.7cm}&&-N^2\int_{-\frac{1}{2}}^{\frac{1}{2}}dx_1dy_1V(u_{12})=
  -\frac{\pi iN^2}{6\sigma\tau}\sum_\pm\left[{\textstyle
  \sum_{I=1}^3B_3\left(\delta_I\!\mp\! u_2\!-\!\frac{\sigma+\tau}{2}\!+\!1\right)
  \!-\!B_3\left(\mp u_2\!-\!\frac{\sigma+\tau}{2}\!+\!1\right)
  }\right]\\
  \hspace*{-.7cm}&&
  =\frac{\pi iN^2}{6\sigma\tau}\sum_\pm\left[\frac{}{}\!
  {\textstyle\sum_I}
  \left({\textstyle (\frac{\sigma+\tau}{2}\pm u_2\!-\!\delta_I\!-\!\frac{1}{2})^3
  -\frac{1}{4}(\frac{\sigma+\tau}{2}\pm u_2\!-\!\delta_I\!-\!\frac{1}{2})}\right)
  {\textstyle -(\frac{\sigma+\tau}{2}\pm u_2\!-\!\frac{1}{2})^3
  +\frac{1}{4}(\frac{\sigma+\tau}{2}\pm u_2\!-\!\frac{1}{2})}\right]
  \nonumber\\
  \hspace*{-.7cm}&&=-\frac{\pi iN^2}{\sigma\tau}\left[{\textstyle
  \delta_1\delta_2\delta_3+\frac{\Delta^3}{12}
  -\frac{\Delta}{4}\left(\frac{1}{3}+2\delta_1\delta_2+2\delta_2\delta_3
  +2\delta_3\delta_1-\delta_1^2-\delta_2^2-\delta_3^2\right)+\Delta u_2^2
  }\right]
  =-\frac{\pi iN^2\delta_1\delta_2\delta_3}{\sigma\tau}\ ,\nonumber
\end{eqnarray}
where $\Delta\equiv \sum_I\delta_I-\sigma-\tau+1$.
We used
\begin{equation}\label{B3-rewrite}
  B_3(z)={\textstyle z^3-\frac{3}{2}z^2+\frac{1}{2}z}
  \ \rightarrow\ B_3(z+1)={\textstyle \left(z+\frac{1}{2}\right)^3
  -\frac{1}{4}\left(z+\frac{1}{2}\right)}\ .
\end{equation}
on the second line, and $\Delta=0$ on the fourth line for the
upper case of (\ref{delta-classify}).

Now one can immediately compute the force, by taking the $u_2$ derivative
of (\ref{potential-constant}). Since the final expression contains no
$u_2$ dependence, one finds that
\begin{equation}
  -\frac{\partial}{\partial u_2}\int dx_1dy_1 V(u_1-u_2)=0\ ,
\end{equation}
proving that our ansatz solves the saddle point equation.
(\ref{potential-constant}) fails to be $u_2$-independent
if one uses the ansatz (\ref{ansatz-epsilon}) with $\epsilon<0$.
We can also compute the saddle point free energy, by integrating
(\ref{potential-constant}) once more in $x_2,y_2$:
\begin{equation}
  \log Z=-N^2\int_{-\frac{1}{2}}^{\frac{1}{2}} dx_2dy_2
  \int_{-\frac{1}{2}}^{\frac{1}{2}} dx_1dy_1 V(u_1-u_2)=
  -\frac{\pi i N^2\delta_1\delta_2\delta_3}{\sigma\tau}\ .
\end{equation}
This again agrees with the free energy of the BPS black holes in
$AdS_5\times S^5$ \cite{Hosseini:2017mds}.

\subsection{A relation to the Bethe ansatz equation}

We would like to provide an interpretation of the vanishing of the force
that we have just proven. We first consider
\begin{equation}
  \frac{\partial}{\partial u_2}\int dx_1 dy_1
  \log \Gamma(z\pm u_{12},\sigma,\tau)=
  \mp\frac{1}{\sigma}\int_{-\frac{1}{2}}^\infty dy
   dx\frac{\partial}{\partial x}
  \left[{\rm Li}_1(e^{2\pi i(u\mp u_2+z)})
  +{\rm Li}_1(e^{2\pi i(u+\sigma+\tau-z-1\pm u_2)})\right]
  \nonumber
\end{equation}
where ${\rm Li}_1$ is defined using analytic continuation
if necessary, and we replaced $\frac{\partial}{\partial u_2}$
by $\mp\frac{1}{\sigma}\frac{\partial}{\partial x}$ and
$\pm\frac{1}{\sigma}\frac{\partial}{\partial x}$ for the
first and the second term. Instead of double-integrating these
to ${\rm Li}_2$ functions, similar to what we did so far in this section,
we integrate with $x$ only to obtain
\begin{equation}
  \frac{\partial}{\partial u_2}\int dx_1 dy_1
  \log \Gamma(z\pm u_{12},\sigma,\tau)=
  \pm\frac{1}{\sigma}\int_{-\frac{1}{2}}^\infty dy
  \left[{\rm Li}_1(e^{2\pi i(y\tau\mp u_2+z-\frac{\sigma}{2})})
  -{\rm Li}_1(e^{2\pi i(y\tau+\sigma+\tau-z-1\pm u_2-\frac{\sigma}{2})})\right]\ .
  \nonumber
\end{equation}
This can be understood as
\begin{equation}\label{theta-log}
  \mp\frac{1}{\sigma}\int_{-\frac{1}{2}}^{\frac{1}{2}}
  dy_1\sum_{n=0}^\infty\left[-\log(1-e^{2\pi i(\pm u_{12}+z-\frac{\sigma}{2}+n\tau)})
  +\log(1-e^{2\pi i(\mp u_{12}+\frac{\sigma}{2}-z-1+(n+1)\tau)})\right]
\end{equation}
where $u_1\equiv y_1\tau$. (This definition of $u_1$ will be assumed below
when accompanied only by single integration $\int dy_1$.)
The $-1$ shift in the exponent of the second log
is again a helpful convention if this log is defined
by the analytic continuation (\ref{log-continue}).
Formally, we can write this as
\begin{eqnarray}\label{gamma-theta-rigorous}
  &&\mp\frac{1}{\sigma}\int_{-\frac{1}{2}}^{\frac{1}{2}}dy_1
  \sum_{n=0}^\infty\log \left[
  (1-e^{2\pi i(z-\frac{\sigma}{2}\pm u_{12})}e^{2\pi in\tau})
  (1-e^{-2\pi i(z+1-\frac{\sigma}{2}\pm u_{12})}e^{2\pi i(n+1)\tau})\right]
  \nonumber\\
  &&\textrm{`}\equiv\textrm{'}\mp\frac{1}{\sigma}\int_{-\frac{1}{2}}^{\frac{1}{2}}
  dy_1 \log\theta({\textstyle z-\frac{\sigma}{2}}\pm u_{12},\tau)\ ,
\end{eqnarray}
where $\theta(z,\tau)$ is the `$q$-theta function' with
$q\equiv e^{2\pi i\tau}$ defined by
\begin{equation}\label{q-theta}
  \theta(z,\tau)\equiv
  \prod_{n=0}^\infty (1-e^{2\pi iz}e^{2\pi in\tau})
  (1-e^{-2\pi i(z+1)}e^{2\pi i(n+1)\tau})
  \sim \prod_{n=0}^\infty (1-e^{2\pi iz}e^{2\pi in\tau})
  (1-e^{-2\pi iz}e^{2\pi i(n+1)\tau})\ .
\end{equation}
(We shall often write it as $\theta(z)$ if no confusions are expected.)
Let us explain the meaning of this calculus. Had one been sloppy about
taking log of functions, an apparently similar result could have been obtained
by a much neater calculation,
\begin{eqnarray}\label{gamma-theta-sloppy}
  &&\frac{\partial}{\partial u_2}\int dx_1 dy_1
  \log \Gamma(z\pm u_{12},\sigma,\tau)=
  -\frac{1}{\sigma}\int_{-\frac{1}{2}}^{\frac{1}{2}} dy_1
  dx_1\frac{\partial}{\partial x_1}
  \log \Gamma(z\pm u_{12},\sigma,\tau )\\
  &&=-\frac{1}{\sigma}\int_{-\frac{1}{2}}^{\frac{1}{2}} dy_1
  \log\frac{\Gamma(z\pm u_{12}\pm\frac{\sigma}{2},\sigma,\tau)}
  {\Gamma(z\pm u_{12}\mp\frac{\sigma}{2},\sigma,\tau)}
  =\mp\frac{1}{\sigma}\int_{-\frac{1}{2}}^{\frac{1}{2}} dy_1
  \log\theta(z-{\textstyle \frac{\sigma}{2}}\pm u_{12},\tau)
  \nonumber
\end{eqnarray}
if one can apply the identity
\begin{equation}\label{Gamma-quasi-period}
  \frac{\Gamma(z+\sigma,\sigma,\tau)}{\Gamma(z,\sigma,\tau)}
  =\theta(z,\tau)
\end{equation}
inside the log. However, it is obscure
what it means to apply an identity inside the log, in particular
if some log functions are defined by analytic continuations.
The precise meaning of the last expression
of (\ref{gamma-theta-sloppy}) is given by the first line of
(\ref{gamma-theta-rigorous}), with continuations (\ref{log-continue})
understood. We have already specified the correct branch
sheet of each log function. With these understood,
we study the force $F$ defined by
\begin{equation}
  F\equiv 2N\int dx_1 dy_1 \partial_{u_2}V(u_1-u_2)=
  \frac{2N}{\sigma}\int dx_1 dy_1 \partial_{x_2}V(u_1-u_2)
  =-\frac{2N}{\sigma}\int dx_1 dy_1 \partial_{x_1}V(u_1-u_2)\ .
\end{equation}
Applying (\ref{gamma-theta-rigorous}), one obtains
\begin{equation}\label{force-collinear}
  F=\frac{N}{\sigma}\int_{-\frac{1}{2}}^{\frac{1}{2}}
  dy_1 \log\left[\frac{\theta(-u_{12}-\frac{\sigma}{2},\tau)}
  {\theta(u_{12}-\frac{\sigma}{2},\tau)}
  \prod_I\frac{\theta(\delta_I+u_{12}-\frac{\sigma}{2},\tau)}
  {\theta(\delta_I-u_{12}-\frac{\sigma}{2},\tau)}\right]
\end{equation}
where $u_1\equiv y_1\tau$,
and all $\log\theta$ functions are understood in the sense of
(\ref{gamma-theta-rigorous}).

Let us study the following function
\begin{equation}\label{f-definition}
  f(u)\equiv\frac{\theta(-u-\frac{\sigma}{2},\tau)}
  {\theta(u-\frac{\sigma}{2},\tau)}\prod_I
  \frac{\theta(\delta_I+u-\frac{\sigma}{2},\tau)}
  {\theta(\delta_I-u-\frac{\sigma}{2},\tau)}
\end{equation}
%\begin{equation}
%  f(u)\equiv\frac{\theta(-u-\sigma,\tau)}
%  {\theta(u,\tau)}\prod_I\frac{\theta(\delta_I+u,\tau)}
%  {\theta(\delta_I-u-\sigma,\tau)}
%\end{equation}
in more detail, which appears inside the log in (\ref{force-collinear}).
We first study
its properties in the usual manner, without worrying about taking the log,
to first get intuitions. We shall then make all the calculations
rigorously inside the log. Using
\begin{equation}\label{theta-shift}
  \theta(z+\tau,\tau)=-e^{-2\pi iz}\theta(z,\tau)\ ,\ \
  \theta(z-\tau,\tau)=-e^{2\pi iz}e^{-2\pi i\tau}\theta(z,\tau)\ ,
\end{equation}
one finds
\begin{eqnarray}\label{f-elliptic}
  f(u+\tau)&=&\frac{-e^{-2\pi i\tau}e^{-2\pi i(u+\frac{\sigma}{2})}
  \theta(-u-\frac{\sigma}{2},\tau)}
  {-e^{-2\pi i (u-\frac{\sigma}{2})}\theta(u-\frac{\sigma}{2},\tau)}
  \prod_I\frac{-e^{-2\pi i(u+\delta_I-\frac{\sigma}{2})}
  \theta(\delta_I+u-\frac{\sigma}{2},\tau)}
  {-e^{-2\pi i\tau}e^{2\pi i(\delta_I-u-\frac{\sigma}{2})}
  \theta(\delta_I-u-\frac{\sigma}{2},\tau)}\nonumber\\
  &=&
  e^{-4\pi i(\sum_I\delta_I-\sigma-\tau)}\cdot
  \frac{\theta(-u-\frac{\sigma}{2},\tau)}{\theta(u-\frac{\sigma}{2},\tau)}
  \prod_I\frac{\theta(\delta_I+u-\frac{\sigma}{2},\tau)}
  {\theta(\delta_I-u-\frac{\sigma}{2},\tau)}=f(u)\ ,
\end{eqnarray}
%\begin{equation}
%  f(u+\tau)=\frac{-e^{-2\pi i\tau}e^{-2\pi i(u+\sigma)}\theta(-u-\sigma,\tau)}
%  {-e^{-2\pi i u}\theta(u,\tau)}
%  \prod_I\frac{-e^{-2\pi i(u+\delta_I)}\theta(\delta_I+u,\tau)}
%  {-e^{-2\pi i\tau}e^{2\pi i(\delta_I-u-\sigma)}
%  \theta(\delta_I-u-\sigma,\tau)}=f(u)\ ,
%\end{equation}
upon using $\sum_I\delta_I-\sigma-\tau\in\mathbb{Z}$.
So the function $f(u_{12})$ appearing in the log in (\ref{force-collinear}) is
double-periodic, $u_{12}+1\sim u_{12}+\tau\sim u_{12}$.
Of course after taking the log, $\log f$ is periodic in both directions
up to $2\pi i\mathbb{Z}$ which we shall clarify shortly.

More specifically, we consider the case with $\sigma=\tau$. Then
the function $f$ can be written as
\begin{eqnarray}\label{f-arrange}
  f(u)&=&\frac{\theta(-u-\frac{\tau}{2},\tau)}{\theta(u-\frac{\tau}{2},\tau)}
  \prod_I\frac{\theta(\delta_I+u-\frac{\tau}{2},\tau)}
  {\theta(\delta_I-u-\frac{\tau}{2},\tau)}\\
  &=&\frac{-e^{2\pi i(\frac{\tau}{2}-u)}e^{-2\pi i\tau}
  \theta(-u+\frac{\tau}{2},\tau)}{\theta(u-\frac{\tau}{2},\tau)}
  \prod_I\frac{\theta(\delta_I+u-\frac{\tau}{2},\tau)}
  {-e^{2\pi i(\frac{\tau}{2}-u+\delta_I)}e^{-2\pi i\tau}
  \theta(\delta_I-u+\frac{\tau}{2},\tau)}\nonumber\\
  &=&e^{2\pi i(2u-\sum_I\delta_I+\tau)}\cdot
  \left(-e^{2\pi i(\frac{\tau}{2}-u)}\right)
  \prod_I\frac{\theta(\delta_I+u-\frac{\tau}{2},\tau)}
  {\theta(\delta_I-u+\frac{\tau}{2},\tau)}
  =-e^{2\pi i(u-\frac{\tau}{2})}
  \prod_I\frac{\theta(\delta_I+u-\frac{\tau}{2},\tau)}
  {\theta(\delta_I-u+\frac{\tau}{2},\tau)}
  \nonumber
\end{eqnarray}
%\begin{eqnarray}
%  f(u)&=&\frac{-e^{2\pi i(z-y\tau)}e^{-2\pi i\tau}\theta(z-y\tau,\tau)}
%  {\theta(y\tau-z,\tau)}\prod_I\frac{\theta(\delta_I+y\tau-z,\tau)}
%  {-e^{2\pi i(\delta_I+z-y\tau)}e^{-2\pi i\tau}\theta(\delta_I+z-y\tau,\tau)}
%  \nonumber\\
%  &=&-e^{2\pi i(y\tau-z)}\prod_I\frac{\theta(\delta_I+y\tau-z)}
%  {\theta(\delta_I+z-y\tau)}
%\end{eqnarray}
where we used (\ref{theta-shift}) on the
second line, $\theta(-z,\tau)=-e^{-2\pi iz}\theta(z,\tau)$
and $\sum_I \delta_I=2\tau$ (mod $\mathbb{Z}$)
on the third line, again without worrying about taking log.
Note that $u-\frac{\tau}{2}$ in the last expression is given by
$u_1-\left(u_2+\frac{\tau}{2}\right)$, where
$u_1=y_1\tau$ is on the segment $(-\frac{\tau}{2},\frac{\tau}{2})$ and
$u_2^\prime\equiv u_2+\frac{\tau}{2}$ is on the segment
$(-\frac{\sigma}{2},\frac{\sigma+2\tau}{2})\rightarrow
(-\frac{\tau}{2},\frac{3\tau}{2})$ at $\sigma=\tau$. If
$u_2^\prime\in (-\frac{\tau}{2},\frac{\tau}{2})$, then $u_2^\prime$
is in the same range as $u_1$. If $u_2^\prime\in(\frac{\tau}{2},\frac{3\tau}{2})$,
we can use the $\tau$ shift invariance of $f(u)$ to replace all $u_2^\prime$
arguments by $u_2^\prime-\tau\in(-\frac{\tau}{2},\frac{\tau}{2})$.
So let us define
\begin{equation}\label{u-bethe}
  u_2^{\rm Bethe}=\left\{\begin{array}{ll}
    u_2^\prime&\textrm{ for }u_2\in(-\frac{\tau}{2},\frac{\tau}{2})\\
    u_2^\prime-\tau&\textrm{ for }u_2\in(\frac{\tau}{2},\frac{3\tau}{2})
  \end{array}\right.
\end{equation}
satisfying $u_2^{\rm Bethe}\in(-\frac{\tau}{2},\frac{\tau}{2})$.
Then one can write
\begin{equation}\label{f-Q}
  f(u_1-u_2)=f(u_1-u_2^{\rm Bethe}{\textstyle +\frac{\tau}{2}})
  =-e^{2\pi i(u_1-u_2^{\rm Bethe})}
  \prod_I\frac{\theta(\delta_I+u_1-u_2^{\rm Bethe},\tau)}
  {\theta(\delta_I-u_1+u_2^{\rm Bethe},\tau)}\ .
\end{equation}
Here we note that the last expression for $f$
is same as the function $Q(u)$ appearing in the Bethe ansatz equation of \cite{Closset:2017bse,Benini:2018mlo}, defined by
\begin{equation}\label{bethe-Q}
  Q(u)=-e^{6\pi i u}
  \frac{\theta(\delta_1-u)\theta(\delta_2-u)\theta(\delta_3-u-2\tau)}
  {\theta(\delta_1+u)\theta(\delta_2+u)\theta(\delta_3+u-2\tau)}
  =-e^{-2\pi i u}\prod_{I=1}^3\frac{\theta(\delta_I-u)}{\theta(\delta_I+u)}
\end{equation}
where we used $\theta(z-2\tau,\tau)=e^{4\pi iz}e^{-6\pi i\tau}\theta(z,\tau)$.
Namely, $f(u_1-u_2^{\rm Bethe}+\frac{\tau}{2})=Q(u_2^{\rm Bethe}-u_1)$.

With these understood, now we review the Bethe ansatz equation.
Suppose we have $N$ variables $u_a$ ($a=1,\cdots,N$) given by
$u_a=-\frac{\tau}{2}+\frac{a\tau}{N}$. In the continuum limit, they are
distributed uniformly on the interval $(-\frac{\tau}{2},\frac{\tau}{2})$.
They satisfy the following Bethe ansatz equation
\begin{equation}\label{BAE}
  1=Q_a(\{u\})\equiv\prod_{b(\neq a)}Q(u_a-u_b)
\end{equation}
with $Q(u)$ given by (\ref{bethe-Q}).
This is the Bethe root of \cite{Hong:2018viz} at $K=1$, $r=0$.
The continuum version of this equation is
\begin{equation}\label{log-Q}
  0=\sum_{b(\neq a)}\log Q(u_a-u_b)\rightarrow
  N\int_{-\frac{1}{2}}^{\frac{1}{2}}dy_1\log Q(u_2^{\rm Bethe}-u_1)
  =N\int_{-\frac{1}{2}}^{\frac{1}{2}}dy_1\log
  f(u_1-u_2^{\rm Bethe}+{\textstyle \frac{\tau}{2}})\ .
\end{equation}
We renamed the continuum variables
$u_b\rightarrow u_1\equiv y_1 \tau$ and
$u_a\rightarrow u_2^{\rm Bethe}$, both in the range
$(-\frac{\tau}{2},\frac{\tau}{2})$. This is precisely the vanishing
condition of the force (\ref{force-collinear}) of our interest.
So we have shown that the saddle point equation at $\sigma=\tau$ is equivalent to
the log of the Bethe ansatz equation within our ansatz. In this viewpoint,
the uniform Bethe root on a segment is obtained by projecting (partially
summing over) the uniform parallelogram distribution along one direction.
What is unclear at this stage
is the $2\pi i\mathbb{Z}$ ambiguities when applying identities inside the log,
but it still establishes the relation solidly. The Bethe
ansatz equation is usually discussed without taking log,
and all the subtleties of $2\pi i\mathbb{Z}$ in our calculus are collected to
the question of what $\log 1$ is on the left hand side of (\ref{BAE}).
In the remaining part of this section, we want to address what this
constant is within our setup.

%%% Saebyeok's definitions %%%

 \def\p{\partial}
 \def\pb{\bar{\partial}}
 \def\a{\alpha}
 \def\b{\beta}
 \def\g{\gamma}
 \def\d{\delta}
 \def\eps{\epsilon}

 \def\th{\theta}
 \def\vt{\th}
 \def\k{\kappa}
 \def\l{\lambda}
 \def\m{\mu}
 \def\n{\nu}
 \def\x{\xi}
 \def\r{\rho}
 \def\u{\upsilon}
 \def\vr{\varrho}
 \def\s{\sigma}
 \def\t{\tau}
 \def\th{\theta}
 \def\z{\zeta }
 \def\vp{\varphi}
 \def\G{\Gamma}
 \def\D{\Delta}
 \def\T{\theta}
 \def\X{\Xi}
 \def\P{\Pi}
 \def\S{\Sigma}
 \def\L{\Lambda}
 \def\O{\Omega}
 \def\o{\omega }
 \def\U{\Upsilon}

\def\beq{\begin{equation}}
\def\eeq{\end{equation}}

It suffices to reconsider all the theta function identities used to establish
(\ref{f-Q}), (\ref{bethe-Q}), rigorously stating their log versions with our
conventions. We have used the $\pm\tau$ shift
identities (\ref{theta-shift}) and the inversion identity
$\theta(-z,\tau)=-e^{-2\pi iz}\theta(z,\tau)$ during the derivation.
Consider the log of the function $f(u_{12})$ defined by (\ref{f-definition}),
where $u_1 = y_1 \t$ and $u_2 = \s x_2 + \t y_2$. Note that
\begin{align} \label{eq:th1}
\begin{split}
    \log \th \left(-u_{12} -\frac{\t}{2},\t \right) &= \sum_{n=0}  ^{\infty} \left( \log (1- e^{2\pi i (-u_{12}-\frac{\t}{2})} e^{2\pi i n \t}) + \log (1- e^{-2\pi i (-u_{12} -\frac{\t}{2} +1 )} e^{2\pi i (n+1) \t} \right) \\
    & = \log (1- e^{2\pi i (-u_{12} -\frac{\t}{2})} ) - \log (1- e^{2\pi i (u_{12}+ \frac{\t}{2} -1)} ) \\
    &\quad+ \sum_{n=0} ^\infty \left( \log (1- e^{2\pi i (-u_{12} +\frac{\t}{2} ) } e^{2\pi i n\t} ) + \log (1- e^{-2\pi i (-u_{12} +\frac{\t}{2} +1) } e^{2\pi i (n+1)\t}) \right) \\
    &= -2\pi i \left( u_{12}+ \frac{\t}{2} \right) + \pi i + \log \th \left(- u_{12} +\frac{\t}{2},\t \right) ,
\end{split}
\end{align}
where one of the two logarithms in the second line is defined by the Taylor series and the other is defined by an analytic continuation. In any of the two cases, we can use the continuation formula which yields the same result given by the last line. In a similar way, we get
\begin{align} \label{eq:th2}
\begin{split}
    \log \th \left(\d_I -u_{12} -\frac{\t}{2},\t \right) &= \sum_{n=0}  ^{\infty} \left( \log (1- e^{2\pi i ( \d_I -u_{12}-\frac{\t}{2})} e^{2\pi i n \t}) + \log (1- e^{-2\pi i (\d_I -u_{12} -\frac{\t}{2} +1 )} e^{2\pi i (n+1) \t} \right) \\
    & = \log (1- e^{2\pi i (\d_I -u_{12} -\frac{\t}{2})} ) - \log (1- e^{2\pi i (-\d_I+ u_{12}+ \frac{\t}{2} -1)} ) \\
    &\quad+ \sum_{n=0} ^\infty \left( \log (1- e^{2\pi i (\d_I -u_{12} +\frac{\t}{2} ) } e^{2\pi i n\t} ) + \log (1- e^{-2\pi i (\d_I -u_{12} +\frac{\t}{2} +1) } e^{2\pi i (n+1)\t}) \right) \\
    &= 2\pi i \left(\d_I - u_{12} - \frac{\t}{2} \right) + \pi i + \log \th \left(\d_I- u_{12} +\frac{\t}{2},\t \right).
\end{split}
\end{align}
Lastly, let us consider
\begin{eqnarray}
    \log \th \left( -u_{12} +\frac{\t}{2},\t \right)
    &=&-\log (1- e^{2\pi i (u_{12} -\frac{\t}{2}-1)}) + \log (1- e^{-2\pi i (u_{12}-\frac{\t}{2}) } ) \\
    &&+\sum_{n=0} ^\infty \log (1- e^{2\pi i (u_{12} -\frac{\t}{2} -1) } e^{2\pi i n \t} ) (1- e^{-2\pi i (u_{12}-\frac{\t}{2})} e^{2\pi i (n+1) \t} )\ .\nonumber
\end{eqnarray}
The first logarithm in the last line has to be defined by analytic continuation when
\begin{equation}
    \text{Im}\left(u_{12} -\frac{\t}{2} -1 + n \t \right) = \text{Im}\,\t \left( n-\frac{1}{2} +y_{12} - x_2 \right) < 0
    \Leftrightarrow  n < x_2 - y_{12} + \frac{1}{2}.
\end{equation}
The number of such non-negative integers $n$ is $M \equiv \text{max}\{\lceil x_2 - y_{12} + \frac{1}{2} \rceil ,0 \}$. Similarly, the second logarithm in the last line is defined by analytic continuation when
\begin{equation}
    \text{Im} \left(-u_{12} + \frac{\t}{2} (n+1)\t \right) = \text{Im}\,\t \left( n+\frac{3}{2} -y_{12} +  x_2 \right)< 0
    \Leftrightarrow n< y_{12} - x_2 -\frac{3}{2}\ .
\end{equation}
There is no such non-negative integer $n$, since the $y_{12}- x_2 < \frac{3}{2}$. Note that
\begin{subequations}
\begin{align}
    &\sum_{n=0} ^\infty \log(1- e^{2\pi i (u_{12} - \frac{\t}{2} -1) } e^{2\pi i n \t}) = -2\pi i M + \sum_{n=0} ^\infty \log (1- e^{ 2\pi i (u_{12} - \frac{\t}{2}) } e^{2\pi i n \t} ) \\
    &\sum_{n=0} ^\infty \log (1- e^{-2\pi i (u_{12}-\frac{\t}{2})} e^{2\pi i (n+1) \t}) =  \sum_{n=0} ^\infty \log (1- e^{-2\pi i (u_{12} -\frac{\t}{2} +1)} e^{2\pi i (n+1) \t} ),
\end{align}
\end{subequations}
by the analytic continuation formula. We can explicitly write the non-negative integer $M$ as
\begin{align}
    M = \begin{cases} \lceil x_2 - y_{12} + \frac{1}{2} \rceil \quad &\text{if} \quad x_2 -y_{12} \geq -\frac{1}{2} \\ \quad\quad\quad 0 &\text{if} \quad -\frac{3}{2} \leq x_2 -y_{12} < -\frac{1}{2}   \end{cases} .
\end{align}
Thus, we can write
\begin{align} \label{eq:th3}
    \log \th\left( -u_{12} +\frac{\t}{2}, \t \right) = -2\pi i \left( u_{12} -\frac{\t}{2} \right) + \pi i - 2\pi i M + \log \th \left( u_{12}-\frac{\t}{2},\t \right)
\end{align}
Combining \eqref{eq:th1}, \eqref{eq:th2}, \eqref{eq:th3}, we obtain
\begin{align}
\begin{split}
    \log f(u_{12}) &= \log \th\left( -u_{12} -\frac{\t}{2} ,\t \right) - \log \th\left( u_{12} -\frac{\t}{2} ,\t \right) \\
    &+\sum_{I=1} ^3 \log \th\left(\d_I + u_{12} -\frac{\t}{2} ,\t \right) - \log \th\left(\d_I -u_{12} -\frac{\t}{2} ,\t \right) \\
    &= 2\pi i \left(u_{12} -\frac{\t}{2} + \frac{1}{2} -M \right) +\sum_{I=1} ^3 \log \left( \d_I +u_{12} -\frac{\t}{2},\t \right) -\log \th \left( \d_I -u_{12} +\frac{\t}{2},\t \right),
\end{split}
\end{align}
where we have used $\sum_{I=1} ^3 \d_I = 2\t -1$ in the second equality.

Now recall the definition of $u_2^\prime\equiv u_2+\frac{\tau}{2}
\in (-\frac{\tau}{2},\frac{3\tau}{2})$
and $u_2^{\rm Bethe}$ (\ref{u-bethe}).
Then one obtains
\begin{align}
\begin{split}
    &\log f(u_{12}) = \log f (u_1 - u_2 ^{\text{Bethe}} + \frac{\t}{2}) \\
    &= 2\pi i \left( u_1 - u_2 ^{\text{Bethe}} +\frac{1}{2} -M \right) + \sum_{I=1} ^3 \log \th\left( \d_I +u_1 -u_2 ^{\text{Bethe}} ,\t \right) - \log \th\left( \d_I -u_1 +u_2 ^{\text{Bethe}} ,\t \right)
\end{split}
\end{align}
when $u_2 ' \in \left(-\frac{\t}{2},\frac{\t}{2} \right)$ and
\begin{align}
\begin{split}
    \log f(u_{12}) &= \log f(u_1 - u_2 ^{\text{Bethe}} + \frac{3\t}{2} ) \\
    & = 2\pi i \left(u_1 -u_2 ^{\text{Bethe}} + \t +\frac{1}{2} -M \right) \\
    &+\sum_{I=1} ^3 \log \th \left(\d_I +u_1 - u_2 ^{\text{Bethe}} + \t ,\t \right) -\log \th \left( \d_I -u_1 + u_2 ^{\text{Bethe}} -\t ,\t \right) \\
    &= 2\pi i \left(u_1 -u_2 ^{\text{Bethe}} - \frac{1}{2} -M \right) + \sum_{I=1} ^3 \log \th\left( \d_I +u_1 - u_2 ^{\text{Bethe}} ,\t \right) - \log \th \left( \d_I -u_1 + u_2 ^{\text{Bethe}} ,\t \right)
\end{split}
\end{align}
when $u_2' \in \left(\frac{\t}{2}, \frac{3\t}{2} \right)$.
All the log functions are defined using our
convention (\ref{theta-log}). This is the precise meaning of the log
of (\ref{f-Q}), where $u_1,u_2^{\rm Bethe}$ are defined to be on the
segment $(-\frac{\tau}{2},\frac{\tau}{2})$.

Finally, again applying (\ref{eq:th2}),
one finds
\begin{eqnarray}
  \hspace*{-1cm}&&\log f(u_{12})=-6\pi i(u_1-u_2^{\rm Bethe})
  +2\pi i\left({\textstyle \frac{1}{2}}-M\right)
  +\log\theta(\delta_1\!+\!u_1\!-\!u_2^{\rm Bethe})
  -\log\theta(\delta_1\!-\!u_1\!+\!u_2^{\rm Bethe})\nonumber\\
  \hspace*{-1cm}&&+\log\theta(\delta_2\!+\!u_1\!-\!u_2^{\rm Bethe})
  -\log\theta(\delta_2\!-\!u_1\!+\!u_2^{\rm Bethe})
  +\log\theta(\delta_3\!+\!u_1\!-\!u_2^{\rm Bethe}\!-\!2\tau)
  -\log\theta(\delta_3\!-\!u_1\!+\!u_2^{\rm Bethe}\!-\!2\tau)
  \nonumber
\end{eqnarray}
when $u_2^\prime\in(-\frac{\tau}{2},\frac{\tau}{2})$, and
\begin{eqnarray}
  \hspace*{-1cm}&&\log f(u_{12})=-6\pi i(u_1-u_2^{\rm Bethe})
  -2\pi i\left({\textstyle \frac{1}{2}}+M\right)
  +\log\theta(\delta_1\!+\!u_1\!-\!u_2^{\rm Bethe})
  -\log\theta(\delta_1\!-\!u_1\!+\!u_2^{\rm Bethe})\nonumber\\
  \hspace*{-1cm}&&+\log\theta(\delta_2\!+\!u_1\!-\!u_2^{\rm Bethe})
  -\log\theta(\delta_2\!-\!u_1\!+\!u_2^{\rm Bethe})
  +\log\theta(\delta_3\!+\!u_1\!-\!u_2^{\rm Bethe}\!-\!2\tau)
  -\log\theta(\delta_3\!-\!u_1\!+\!u_2^{\rm Bethe}\!-\!2\tau)
  \nonumber
\end{eqnarray}
when $u_2^\prime\in (\frac{\tau}{2},\frac{3\tau}{2})$. These are
what we precisely mean by $\log Q(u_2^{\rm Bethe}-u_1)$ in (\ref{log-Q}).

Note that the significance of the $\epsilon$ deformation (\ref{ansatz-epsilon})
became obscure in our final expression for the Bethe ansatz equation.
For instance, had one chosen the wrong sign for $\epsilon<0$, one would have
got $-\pi i$ on the last line of (\ref{eq:th1}) instead of $+\pi i$.
This would have affected the final expression
for $\log f$ only by an extra constant of the form $2\pi i\mathbb{Z}$, whose
exponentiation yields the same Bethe equation. So we find that
the map of the saddle point to the Bethe root is at best one-sided, e.g.
the Bethe roots being unable to detect the sign of $\epsilon$.
Recall that this deformation
was needed because otherwise there are eigenvalues $u_1$ and $u_2$ which differ by
$u_{12}=\tau\mathbb{Z}\neq 0$ that makes some gaugino operators massless.
This was explained in the paragraph containing (\ref{gaugino-fugacity}).
To repeat the explanation at $\sigma=\tau$, the eigenvalues lie on an interval
$(-\tau,\tau)$. So there exists a pair $u_1$ to any $u_2$
which hits one of the singularities $u_{12}=\pm\tau$ which demands
regularized definition of the ansatz.
On the other hand, the Bethe root obtained (in our viewpoint) by
projecting the parallelogram along $x$ direction is distributed on
the reduced interval $(-\frac{\tau}{2},\frac{\tau}{2})$, causing no
divergence problems. So it is natural that the Bethe
ansatz equation does not detect the subtle details of the true saddle point
such as the sign of $\epsilon$.

Similarly, for a more general case of collinear $\sigma$ and $\tau$ with $\sigma/\tau \in \mathbb{Q}$, 
the relation between our ansatz and the Bethe ansatz equation studied in \cite{Benini:2020gjh} can be shown as follows.
Since $\sigma/\tau \in \mathbb{Q}$, one can define $\sigma\equiv a\omega, \tau \equiv b\omega$ where $(a, b)$ are co-prime integers.
The force equation (\ref{force-collinear}) can be rewritten as follows.
\begin{equation}
    F\propto \sum_{i=0}^{b-1}\int_{0}^{1} dy_1 \log\left[\frac{\theta(u_2 +i\omega+\omega y_1-\frac{\sigma+\tau}{2},b\omega)}{\theta(i\omega+\omega y_1-u_2-\frac{\sigma+\tau}{2},b\omega)}\prod_{I=1}^{3}\frac{\theta(\delta_I +i\omega+\omega y_1-u_2-\frac{\sigma+\tau}{2},b\omega)}{\theta(\delta_I+u_2 +i\omega+\omega y_1-\frac{\sigma+\tau}{2},b\omega)}\right].
\end{equation}
Therefore,
\begin{equation}
    F\propto \int_{0}^{1} dy_1 \log\left[\frac{\theta(u_2 - \omega y_1+\omega-\frac{\sigma+\tau}{2},\omega)}{\theta(\omega y_1-u_2-\frac{\sigma+\tau}{2},\omega)}\prod_{I=1}^{3}\frac{\theta(\delta_I +\omega y_1-u_2-\frac{\sigma+\tau}{2},\omega)}{\theta(\delta_I+u_2 -\omega y_1+\omega-\frac{\sigma+\tau}{2},\omega)}\right],
\end{equation}
where we use $\prod_{i=0}^{b-1}\theta(x+i\omega,b\omega)=\theta(x,\omega)$.
We further define $n_1, n_2$ and $x$ as follows
\begin{equation}
    \frac{\sigma+\tau}{2}-\omega-u_2 = (n_1-x)\omega, \quad
    \frac{\sigma+\tau}{2}+u_2 = (n_2+x)\omega.
\end{equation}
Note that $n_1$ and $n_2$ are integers and $x \in [0,1)$.
Finally, let $\omega y_1-\omega x = u_{12}'$, so that
\begin{equation}
    F\propto \int_{0}^{1} dy_1 \log\left[\frac{\theta(u_{21}'-n_1\omega,\omega)}{\theta(u_{12}'-n_2\omega,\omega)}\prod_{I=1}^{3}\frac{\theta(\delta_I +u_{12}'-n_2\omega,\omega)}{\theta(\delta_I+u_{21}'-n_1\omega,\omega)}\right]
\end{equation}
Using equation (3.38), one can show that 
\begin{equation}
    F\propto \int_{0}^{1} dy_1 \log\left[-e^{-6\pi i u_{12}'}\frac{\theta(\delta_1+u_{12}',\omega)}{\theta(\delta_1-u_{12}',\omega)}\frac{\theta(\delta_2+u_{12}',\omega)}{\theta(\delta_2-u_{12}',\omega)}\frac{\theta(-\delta_1-\delta_2+u_{12}',\omega)}{\theta(-\delta_1-\delta_2-u_{12}',\omega)}\right]=0,
\end{equation}
which is the continuum limit of the Bethe ansatz equation.

Note that the `off-shell' integrand of our matrix model and that of the Bethe ansatz approach are different.
In this subsection, we found a relation between their on-shell conditions, at least for particular saddle points. 
In this sense, our approach is also related to the elliptic extension approach of \cite{Cabo-Bizet:2019eaf, Colombo:2021kbb}. 
In \cite{Cabo-Bizet:2019eaf}, the on-shell physical quantities of their saddle points were shown to agree with those of the corresponding Bethe roots.
So in a broader sense, all three approaches yield the same on-shell results.
Also, the elliptic extension approach is similar to our approach in that both use the periodic nature of the integrand.

\section{Multi-cut saddle points}

In this section, we construct multi-cut saddle points. For a technical reason,
we only consider the case with collinear $\sigma,\tau$.\footnote{
For non-collinear $\sigma,\tau$, they do not satisfy the saddle point
equation. Also, we could not find a modification of the saddle point problem like
section 2.1 which makes them saddle points. Unlike the single-parallelogram ansatz,
we find that after $SL(3,\mathbb{Z})$ modular transformation there are extra poles
included in the parallelogram as well as the zeros from the Haar measure, causing
more complications. We feel that this is related to the gravitational stability
issue of \cite{Aharony:2021zkr}.} Our $K$-cut ansatz is roughly given by
\begin{equation}\label{K-cut}
  u_A(x,y)\equiv \frac{A}{K}+\sigma x+\tau y
\end{equation}
with $x,y\in(-\frac{1}{2},\frac{1}{2})$,
and $A=0,1,\cdots, K-1$ labels the $K$ groups of
eigenvalues forming $K$ cuts. The `$\epsilon$ deformations' of this
ansatz will be specified below, depending on the values of chemical
potentials. Although these are linear
cuts, we again make a 2-parameter labelling of eigenvalues with
uniform 2d distributions.
Each cut contains $\frac{N}{K}$ eigenvalues, at equal filling fraction,
so the density function is given by $\rho_A(x,y)=\frac{1}{K}$.

The large $N$ free energy $\log Z$ of (\ref{K-cut})
in the continuum limit is given by
\begin{equation}
  -\frac{N^2}{K}\int_{-\frac{1}{2}}^{\frac{1}{2}} dx_1dy_1
  \int_{-\frac{1}{2}}^{\frac{1}{2}} dx_2dy_2
  \sum_{A=0}^{K-1}V(u_{12}+{\textstyle \frac{A}{K}})\equiv
  -\frac{N^2}{K}\int_{-\frac{1}{2}}^{\frac{1}{2}} dx_1dy_1
  \int_{-\frac{1}{2}}^{\frac{1}{2}} dx_2dy_2 V_K(u_{12})\ ,
\end{equation}
where $V$ is given by (\ref{potential}), and
$V_K(u)\equiv \sum_{A=0}^{K-1}V(u+\frac{A}{K})$.
The force on the eigenvalue at $u_2$ in the $A=0$ cut is given by
\begin{equation}\label{force-K=2}
  F=2\frac{N}{K}\int dx_1 dy_1\frac{\partial}{\partial u_2}
  V_K(u_{12})
\end{equation}
after plugging in our ansatz.
If this $F$ is zero, the forces on eigenvalues in the
other cuts also vanish by cyclicity. $F=0$ can be shown by confirming that
\begin{equation}\label{potential-function-K=2}
  \int dx_1 dy_1 V_K(u_{12})
\end{equation}
is independent of $u_2$.
We start by noting that
$V_K(u)$ can be written in terms of
\begin{equation}
 \log \Gamma(z,\sigma,\tau)+\Gamma(z+{\textstyle \frac{1}{K}},\sigma,\tau)
 +\cdots+\log\Gamma(z+{\textstyle \frac{K-1}{K}})
 =\log\Gamma(Kz,K\sigma,K\tau)\ .
\end{equation}
So the calculation of
(\ref{potential-function-K=2}) can be done in a manner similar to the
case with $K=1$, by replacing parameters by $K$ times them.
Here we have to be careful about the ranges of the imaginary parts of
$\delta_I,\sigma,\tau$, since multiplying them by $K$ takes them away
from our convention (\ref{delta-classify}).
To be definite, we consider $\delta_I$'s in the
upper case of (\ref{delta-classify}), satisfying
$\sum_{I=1}^3\delta_I=\sigma+\tau-1$. Recall that such $\delta_I$'s
were parametrized as
$\delta_I=-a_I+b_I(\sigma+\tau)$, with
$0<a_I<1$, $0<b_I<1$ satisfying $\sum_Ia_I=\sum_I b_I=1$.
Then $K\delta_I$'s and
$K\sigma$, $K\tau$ satisfy
\begin{equation}
  K\delta_I=K\sigma+K\tau-K\ .
\end{equation}
Here we define $\{Ka_I\}\equiv Ka_I-\lfloor Ka_I\rfloor$, which
measures the fractional part of $Ka_I\in [0,1)$.
Then one finds $\sum_{I=1}^3\{Ka_I\}\in [0,3)$, which has to be
an integer since $\sum_I a_I=1$.
The case with $\sum_{I=1}^3\{Ka_I\}=0$ is very exceptional,
which can be met only if all three $Ka_I$ are integers (because we should
have $\{Ka_I\}=0$ for all $I$'s). This is possible for fine-tuned choices
of $a_I$'s, e.g. at $a_1=a_2=a_3=\frac{1}{3}$ and $K=3$ or
$a_1=a_2=\frac{1}{4}$, $a_3=\frac{1}{2}$ and $K=4$, etc. We shall understand
these special cases with small deformations, so that they satisfy either
$\sum_{I=1}^3\{Ka_I\}=1$ or $2$. For instance, for $a_1=a_2=a_3=\frac{1}{3}$
and $K=3$, slightly reducing $a_1,a_2$ and slightly increasing $a_3$ will
make $\{3a_1\}=\{3a_2\}\lessapprox 1$ and $0<\{3a_3\}\ll 1$, making
them satisfy $\sum_I\{3a_I\}=2$. (Slightly reducing $a_1$ and slightly
increasing $a_2,a_3$ will yield $\sum_{I}\{3a_I\}=1$.) With these understood,
we define new parameters as
$\sigma^\prime\equiv K\sigma$, $\tau^\prime\equiv K\tau$ and
\begin{equation}\label{delta-prime}
  \delta_I^\prime=
  \left\{
  \begin{array}{ll}
    -\{Ka_I\}+b_I(\sigma^\prime+\tau^\prime)&\textrm{if }
    \sum_{I=1}^3\{Ka_I\}=1\\
    1-\{Ka_I\}+b_I(\sigma^\prime+\tau^\prime)&\textrm{if }
    \sum_{I=1}^3\{Ka_I\}=2
  \end{array}
  \right.\ .
\end{equation}
In the upper and lower cases, $\delta_I^\prime$ and $\sigma^\prime$,
$\tau^\prime$ satisfy
$\sum_I\delta_I^\prime-\sigma^\prime-\tau^\prime=\mp 1$, respectively.

To compute (\ref{potential-function-K=2}), we should compute
\begin{equation}\label{log-gamma-K}
  \int_{-\frac{1}{2}}^{\frac{1}{2}}
  dx_1 dy_1\log \Gamma(Kz\pm Ku_{12},K\sigma,K\tau)
\end{equation}
at $z=\delta_I$ or $z=0$,
which can again be done by integrating over two log functions
over $x,y\in(-\frac{1}{2},\infty)$ after a manipulation similar to
(\ref{log-gamma}). In fact one can recycle the calculation at $K=1$
as follows. At the infinity branch where log functions can be defined
by Taylor expansion, we regard $K\delta_I$ as $\delta_I^\prime$ by
trivial period shifts valid at infinity. So the parameters
$\delta_I^\prime,\sigma^\prime,\tau^\prime$ belong to one of the
two cases of (\ref{delta-classify}). Also, in the ansatz for the
first cut $A=0$, $u^\prime\equiv Ku$ given in terms of
$\sigma^\prime,\tau^\prime$ takes precisely the same form as the
single-cut configuration. Therefore, the calculations for the single-cut
can be literally repeated here. As for the three terms at $z=\delta_I$,
the continuous extensions of log functions can be made similarly,
using (\ref{log-continue}). As for the terms at $z=0$, we need to
refine the ansatz with $\epsilon$ like (\ref{ansatz-epsilon}), depending
on which condition of (\ref{delta-classify}) is met by $\delta_I^\prime$.
In the upper case of (\ref{delta-classify}) (or (\ref{delta-prime})),
the ansatz deformation is precisely given in the same direction
as (\ref{ansatz-epsilon}), i.e.
\begin{equation}
  u_A(x,y)=\frac{A}{K}+e^{-i\epsilon}(\sigma x+\tau y)\ \ \ ,
  \ \ \ \epsilon>0\ .
\end{equation}
This deformation yields $u_2$-independent (\ref{potential-function-K=2}),
implying $F=0$.
In the lower case of (\ref{delta-classify}) or (\ref{delta-prime}),
this is the `conjugate sector'  so
the ansatz deformation guaranteeing $F=0$ is given by
\begin{equation}
  u_A(x,y)=\frac{A}{K}+e^{+i\epsilon}(\sigma x+\tau y)\ .
\end{equation}
In both cases, the integral (\ref{potential-function-K=2})
is given by
\begin{equation}\label{potential-constant-K}
  -\frac{N^2}{K}
  \int _{-\frac{1}{2}}^{\frac{1}{2}}dx_1 dy_1 V_K(u_{12})=
  -\frac{\pi i N^2\delta_1^\prime\delta_2^\prime\delta_3^\prime}
  {K\sigma^\prime\tau^\prime}
\end{equation}
by repeating (\ref{potential-constant}). This is independent of $u_2$,
which proves $F=0$.
Integrating (\ref{potential-constant-K}) once more in $x_2,y_2$, one
obtains the following large $N$ free energy
\begin{equation}\label{free-K-prime}
  \log Z=-\frac{\pi iN^2\delta_1^\prime\delta_2^\prime\delta_3^\prime}
  {K\sigma^\prime\tau^\prime}\ .
\end{equation}
Inserting the values of primed variables, one obtains
\begin{equation}
  \log Z=\left\{
  \begin{array}{ll}
    -\frac{\pi iN^2\prod_{I=1}^3
    \left(\delta_I+\frac{\lfloor Ka_I\rfloor}{K}\right)}{\sigma\tau}
    &\textrm{ for the upper case of (\ref{delta-prime})}
    \\
    -\frac{\pi iN^2\prod_{I=1}^3
    \left(\delta_I+\frac{\lfloor Ka_I\rfloor}{K}+\frac{1}{K}\right)}{\sigma\tau}
    &\textrm{ for the lower case of (\ref{delta-prime})}
  \end{array}\right.
  \ .
\end{equation}
Note again that $\delta_I$'s are in the sector defined by the
upper signs of (\ref{delta-classify}). Also, the cases with
all $Ka_I$'s being integral should be understood with care, by
slightly moving them away from the integral values as
illustrated above (\ref{delta-prime}).

The entropy can be obtained by Legendre transforming the
entropy function at fixed charges:
\begin{equation}
  S(Q_I,J_i;\delta_I,\sigma,\tau)=
  \log Z(\delta_I,\sigma,\tau)
  -2\pi i\sum_I\delta_IQ_I-2\pi i\sigma J_1
  -2\pi i\tau J_2\ .
\end{equation}
After extremizing this function with $\delta_I,\sigma,\tau$ subject
to the constraint $\sum_I\delta_I-\sigma-\tau=-1$, one takes the real part
${\rm Re}(S)$ to get the entropy \cite{Choi:2018hmj,Choi:2018vbz,Agarwal:2020zwm}.
This computation can be done easily using
the universal form (\ref{free-K-prime}), by noting that
primed variables $\delta_I^\prime$, $\sigma^\prime$, $\tau^\prime$ are
$K$ times $\delta_I$, $\sigma$, $\tau$ apart from real constant shifts.
Namely, the entropy function is given by
\begin{equation}
  S=-\frac{\pi iN^2\delta_1^\prime\delta_2^\prime\delta_3^\prime}
  {K\sigma^\prime\tau^\prime}
  -2\pi i\sum_I\left(\delta_I^\prime+\cdots\right)\frac{Q_I}{K}
  -2\pi i\sigma^\prime\frac{J_1}{K}
  -2\pi i\tau^\prime\frac{J_2}{K}\ ,
\end{equation}
where $\cdots$ are real constants which only affect $S$ by an imaginary
constant. To compute ${\rm Re}(S)$, we can ignore them.
So without these terms, the Legendre transformation takes completely
same form as that at $K=1$, with replacements
$N^2,Q_I,J_i\rightarrow\frac{N^2}{K},\frac{Q_I}{K},\frac{J_i}{K}$.
The entropy at $K=1$ is homogeneous degree $1$ in scaling $N^2$ and all the
other extensive quantities with a same factor (which is a basic property of
AdS$_5$ black holes). Therefore, the entropy of the $K$-cut saddles are related
to that of the basic saddle at $K=1$ by
$S_K(Q_I,J_i)=\frac{1}{K}S_{K=1}(Q_I,J_i)$. So they would be subdominant
saddle points, compared to the basic saddle at $K=1$, of the Euclidean quantum
gravity. \cite{Aharony:2021zkr} suggested the gravity duals of these saddles
at general $K$ as $\mathbb{Z}_K$ quotients
of the analytically continued Euclidean saddle for the Lorentzian black hole
at $K=1$.

When $\sigma=\tau$,
the final results for $\log Z$ are same as the $K$-cut Bethe roots of
\cite{Hong:2018viz} at $r=0$, although the eigenvalue configurations are
different. Our linear density function is triangular on each cut, which
is a segment $(-\tau,\tau)$. On the other hand, each cut in the Bethe
root is a uniform distribution along a segment $(-\frac{\tau}{2},\frac{\tau}{2})$.
Similar to our analysis of section 3, we expect that integrating
the force function $F$ over either $x$ or $y$ first will project our
saddle point equation to the Bethe ansatz equation. We shall not study
the details here.

We comment that when the complex chemical potentials are
in a particular regime,
we find a one parameter generalization of the $K$-cut solution
when $K\equiv 2m$ is even.
To understand this, let us again start from the following
$\epsilon$-deformed ansatz with a free parameter $\nu\in[0,1]$
\begin{equation}\label{ansatz-epsilon-K}
  u(x,y)=\left\{
  \begin{array}{llll}
    \frac{A}{K}+e^{-i\epsilon}(\sigma x+\tau y)&,&
    \rho(x)=\frac{2\nu}{K}&\textrm{if }A=\textrm{even}\\
    \frac{A}{K}+e^{-i\epsilon}(\sigma x+\tau y)&,&
    \rho(x)=\frac{2(1-\nu)}{K}&\textrm{if }A=\textrm{odd}
  \end{array}
  \right.\ .
\end{equation}
Namely, there are $\frac{2N\nu}{K}$ eigenvalues in each cut at
even $A=0,2,\cdots,2m-2$, and $\frac{2N(1-\nu)}{K}$ eigenvalues
at odd $A=1,3,\cdots, 2m-1$.
The force is given by the $u_2$ derivative of
\begin{eqnarray}\label{potential-function-K-nu}
  &&\int_{-\frac{1}{2}}^{\frac{1}{2}} dx_1 dy_1
  \sum_{l=0}^{m-1}\left[\nu V({\textstyle u_{12}+\frac{2l}{K}})
  +(1-\nu)V({\textstyle u_{12}+\frac{2l+1}{K}})\right]\nonumber\\
  &&=\int dx_1 dy_1
  \left[(1-\nu)\sum_{A=0}^{K-1}V({\textstyle u_{12}+\frac{A}{K}})+
  (1-2\nu)\sum_{l=0}^{m-1}V(\textstyle {u_{12}+\frac{l}{m}})\right]\nonumber\\
  &&=\int dx_1 dy_1
  \left[(1-\nu)V_K(u_{12})+(1-2\nu)V_m(u_{12})\right]
\end{eqnarray}
when $u_2$ is on the $A=0$ cut. If this vanishes for arbitrary
$\nu\in[0,1]$, then the forces acting on eigenvalues on different cuts
also vanish. Now repeating the discussions below (\ref{log-gamma-K}),
one finds that the $\epsilon$-deformed ansatz (\ref{ansatz-epsilon-K})
makes $\int dx_1 dy_1 V_K(u_{12})$ to
be $u_2$-independent if $\epsilon>0$ and $\sum_I\{Ka_I\}=1$,
or if $\epsilon<0$ and $\sum_I\{Ka_I\}=2$. Similarly,
$\int dx_1 dy_1 V_m(u_{12})$ would be $u_2$-independent
if $\epsilon>0$ and $\sum_I\{m a_I\}=1$,
or $\epsilon<0$ and $\sum_I\{m a_I\}=2$. Therefore,
if $\sum_I\{Ka_I\}$ and $\sum_I\{ma_I\}=\sum_I\{\frac{K}{2}a_I\}$
have same value between $1$ and $2$, one can set the sign of $\epsilon$
so that both terms on the last line of (\ref{potential-function-K-nu})
separately vanish. So in this case, we have constructed a saddle point
which admits a nontrivial filling fraction of eigenvalues.
The free energy is given by
\begin{equation}
  \log Z=
    -\frac{\pi iN^2}{\sigma\tau}
    \left[(1-2\nu)^2\prod_{I=1}^3
    \left(\delta_I+{\textstyle \frac{\lfloor ma_I\rfloor}{m}}\right)
    +4\nu(1-\nu)\prod_{I=1}^3\left(\delta_I+{\textstyle
    \frac{\lfloor Ka_I\rfloor}{K}}\right)\right]
\end{equation}
if $\sum_I\{Ka_I\}=\sum_{I}\{ma_I\}=1$, and
\begin{equation}
  \log Z=
    -\frac{\pi iN^2}{\sigma\tau}
    \left[(1-2\nu)^2\prod_{I=1}^3
    \left(\delta_I+{\textstyle \frac{\lfloor ma_I\rfloor}{m}+\frac{1}{m}}
    \right)+4\nu(1-\nu)
    \prod_{I=1}^3\left(\delta_I+{\textstyle
    \frac{\lfloor Ka_I\rfloor}{K}+\frac{1}{K}}\right)\right]
\end{equation}
if $\sum_I\{Ka_I\}=\sum_{I}\{ma_I\}=2$.

To be concrete, we consider the case with $K=2$. $a_I$'s would satisfy
the condition $\sum_I\{2a_I\}=\sum_{I}\{a_I\}=1$ if
$a_1,a_2\in(0,\frac{1}{2})$ and $a_3\in(\frac{1}{2},1)$.
The fundamental domain of this solution is $\nu\in[0,\frac{1}{2}]$
since it has a symmetry under $\nu\rightarrow 1-\nu$ combined with
an overall shift of $u_a\rightarrow u_a+\frac{1}{2}$.
It continuously interpolates the one-cut solution $(K,r)=(1,0)$ at
$\nu=0$ and the two-cut solution $(K,r)=(2,0)$ at $\nu=\frac{1}{2}$.
The free energy is given by
\begin{equation}
  \log Z=-\frac{\pi iN^2}{\sigma\tau}
    \left[(1-2\nu)^2\delta_1\delta_2\delta_3
    +4\nu(1-\nu)\delta_1\delta_2(\delta_3
    +{\textstyle \frac{1}{2}})\right]
    =-\frac{\pi iN^2\delta_1\delta_2\left(\delta_3+2\nu(1-\nu)\right)}
    {\sigma\tau}\ .
\end{equation}
Its Legendre transformation at fixed charges $Q_I$, $J_i$ can be
easily done by noting that
\begin{equation}
  \hspace*{-0.5cm}
  \log Z-2\pi i\sum_I \delta_I Q_I-2\pi i\sigma J_1-2\pi i\tau J_2
  \sim(1-2\nu+2\nu^2)\left[
  -\frac{\pi iN^2\hat\delta_1\hat\delta_2\hat\delta_3}
  {\hat\sigma\hat\tau}
  -2\pi i\sum_I\hat\delta_I Q_I-2\pi i\hat\sigma J_1
  -2\pi i\hat\tau J_2\right]
\end{equation}
where $\sim$ holds up to an irrelevant
imaginary constant, and
$\hat{\delta}_I$, $\hat\sigma$, $\hat\tau$ defined by
\begin{equation}
  \hat\delta_{1,2}\equiv\frac{\delta_{1,2}}{1-2\nu+2\nu^2}\ ,\ \
  \hat\delta_3\equiv\frac{\delta_3+2\nu(1-\nu)}{1-2\nu+2\nu^2}\ ,\ \
  \hat\sigma\equiv\frac{\sigma}{1-2\nu+2\nu^2}\ ,\ \
  \hat\tau\equiv\frac{\tau}{1-2\nu+2\nu^2}
\end{equation}
satisfy $\sum_I\hat\delta_I-\hat\sigma-\hat\tau=-1$.
The extremization of the expression inside the square bracket
is completely the same as the free energy at $K=1$.
So taking the real part of the extremized entropy
function, the entropy of our new saddle point at filling fraction
$\nu$ is given by
\begin{equation}
  {\rm Re}S_{K=2,\nu}(Q_I,J_i)=(1-2\nu+2\nu^2){\rm Re}S_1(Q_I,J_i)\ .
\end{equation}
It will be interesting to seek for the gravity duals of these solutions,
for instance in Euclidean quantum gravity by
generalizing \cite{Aharony:2021zkr}.

\section{Conclusion}

In this paper, we found exact large $N$ saddle points of the
$\mathcal{N}=4$ index which are dual to BPS black holes in $AdS_5\times S^5$.
We employed two different approaches. Firstly, when the complex
chemical potentials $\sigma,\tau$ for the two angular momenta $J_1,J_2$
in AdS are non-collinear, we showed that a novel areal distribution of
eigenvalues illustrated in Fig. \ref{parallel-cartoon} solves the saddle point
equation defined after applying the integral identity (\ref{haar-identity}).
$SL(3,\mathbb{Z})$ modularity of the elliptic gamma function was used to prove this.
Secondly, when $\sigma,\tau$ are collinear, we showed that linear distributions
obtained by collapsing the areal distribution also solves
the traditional saddle point equation. The saddle points we constructed precisely
account for the entropies of the dual black holes \cite{Kunduri:2006ek}.
In the remaining part of this section, we emphasize several subtle
structures of our results, and also suggest possible future directions.

In the collinear case, we found that an `$i\epsilon$ type' deformation
is needed to precisely define our large $N$ saddle point
ansatz. We interpreted that $\epsilon$ is related to small $\frac{1}{N}$.
Without such a refined definition, the continuum
eigenvalue distribution hits the singularity of the potential.
Unlike the principal-valued integral which excludes the unphysical
self-interaction from the Haar measure potential, this singularity
comes from interactions of distinct eigenvalues so should be
avoided in any sensible large $N$ ansatz. In our leading large $N$ calculus,
only the sign of $\epsilon$ mattered.

More physically, these singularities very close to our
saddle point configurations are where the matrix elements of the
gaugino operators become massless. For some eigenvalue pairs,
these operators have `effective fugacities' greater than $1$
which means these operators may condense. More generally, whenever
we made analytic continuations of log functions in section 3, using
(\ref{log-continue}), the corresponding operators could have condensed.
Understanding their structures may shed more light on more general
types of black holes, for instance related to the hairy AdS black holes
where certain operators assume nonzero expectation values in the dual
CFT. For instance, hairy black holes in $AdS_5$ and
$AdS_5\times S^5$ were constructed
\cite{Basu:2010uz,Bhattacharyya:2010yg,Markeviciute:2018yal,Markeviciute:2018cqs}.
Also, it will be interesting to see a more direct connection between
our light gaugino operators and the light near-horizon modes on
the BPS black holes.

Perhaps as a related matter, we also discuss the integration contour
and poles/residues. The subtleties summarized
in the previous two paragraphs appear because the matrix integral contour
is analytically continued. Since our final saddle point has many pairs of
eigenvalues whose log potentials require analytic continuations
beyond their radii of convergence, it is quite likely that the full
contour deformation would cross the poles (bosonic branch points)
of the integrand. When the contour crosses a pole,
various terms can appear. The first term is the full
$N$ dimensional contour integral, where the contour passed through the pole.
This is the term that we studied in this paper. On the other hand,
when a contour crosses a pole, one also finds an extra term from
its residue. One may replace $n$ ($<N$) of the $N$ integral
variables by their pole values, obtaining a term of the form
\begin{equation}\label{contour-residue}
  (\textrm{residue of } n \textrm{ dimensional integral})
  \times (N-n \textrm{ dimensional contour integral})\ .
\end{equation}
We are tempted to interpret
the first factor as the $n$ dual giant gravitons
\cite{McGreevy:2000cw,Grisaru:2000zn}
formed outside the event horizon of a core black hole made of $N-n$
eigenvalues. This interpretation sounds heuristic to us for the
following reasons. There are two types of giant gravitons
\cite{McGreevy:2000cw,Grisaru:2000zn}, which are D3-branes
wrapping contractible $S^3$ cycles in either $AdS_5$ or $S^5$.
The dual giant gravitons are pointlike in $S^5$, while
occupying a spatial $S^3$ in $AdS_5$ at a fixed AdS radius.
They are domain walls in $AdS_5$, reducing the RR 5-form flux
inside it by $1$ unit. So inside $n$ dual giant gravitons,
the core black hole will feel only $N-n$ units of RR-flux.
Since the second factor of (\ref{contour-residue})
takes the form of rank $N-n$ integral, it would yield a black hole like
saddle point in certain $U(N-n)$ gauge theory, qualitatively agreeing
well with the bulk picture. Also note that the first factor has
definite values of $n$ eigenvalues, which are often viewed as the
radial locations of dual giant gravitons in AdS \cite{Kinney:2005ej,Mandal:2006tk}.
Finally, there are studies on the color superconductivity
using these branes in the gravity dual \cite{Henriksson:2019zph},
which in the BPS sector should necessarily include the hairs carrying
other global charges.

Even if this picture is correct, such configurations look somewhat
different from the hairy black holes of
\cite{Basu:2010uz,Bhattacharyya:2010yg,Markeviciute:2018yal,Markeviciute:2018cqs}
constructed with the condensation of the Kaluza-Klein gravitons within the
gravity approximation. In the vacuum $AdS_5$,
shrinking the dual giant graviton by reducing its energy makes the $S^3$
small, converging to the point-like graviton picture when the energy is small.
The giant graviton expanded in $S^5$ can also shrink to the same point-like
graviton, and the two brane descriptions provide
complementary descriptions of the $\frac{1}{8}$-BPS sector
\cite{Biswas:2006tj,Mandal:2006tk}. Once there is a core black hole at
the center of AdS, giant gravitons are still contractible
in $S^5$. But a dual giant graviton in this black hole background
can `shrink' only until its radial position reaches
the event horizon. In fact we have made a provisional study of both
types of giant graviton probes in the background of
\cite{Gutowski:2004ez}. The behaviors of giant gravitons in $S^5$ are
well connected to the point-like gravitons, and appear to
exhibit features somewhat similar to the BPS hairy black holes reported
in \cite{Markeviciute:2018cqs} constructed using the KK graviton modes.
The dual giant graviton probes are somewhat trickier for us to interpret.
In any case, we think there are many interesting questions
in this direction.

In section 4, we constructed multi-cut solutions, whose physics is the same as
the multi-cut Bethe roots of \cite{Hong:2018viz}. We also provided
further generalizations of these multi-cut solutions with
nontrivial filling fractions on the cuts. Within our ansatz, such
generalized filling fractions were allowed only when the chemical potentials
$\delta_I$ are in a particular regime. It will be interesting to find
their gravity duals.

We also note that some of the large $N$ techniques explored in this paper
might find applications to study black holes in AdS$_4$/CFT$_3$,
AdS$_6$/CFT$_5$ or AdS$_7$/CFT$_6$. These problems have been
studied in \cite{Choi:2019zpz,Nian:2019pxj,Choi:2019dfu},
\cite{Choi:2019miv,Crichigno:2020ouj} and
\cite{Choi:2018hmj,Nahmgoong:2019hko,Lee:2020rns}, but we think
we can do more interesting large $N$ studies.

We finally remark that the treatment of our section 2, applying the identity
(\ref{haar-identity})
to slightly change the saddle point problem, might find useful applications
in other matrix models. Although this technique is familiar in
some branches of our community (e.g. enumerating BPS operators more efficiently
via contour integral), we are not aware of this idea applied to construct
large $N$ saddle points. The solutions we got after this procedure also
look quite novel, in that we found areal distributions rather than
traditional linear cut distributions. This approach might be helpful
in other matrix model problems.

\vskip 0.5cm

\hspace*{-0.8cm} {\bf\large Acknowledgements}
\vskip 0.2cm

\hspace*{-0.75cm}
We thank Dongmin Gang, Kimyeong Lee, June Nahmgoong and Jaewon Song
for helpful discussions. This work is supported by a KIAS Individual Grant
(PG081601) at Korea Institute for Advanced Study (SC),
the US Department of Energy under grant DE-SC0010008 (SJ) and
the National Research Foundation of Korea (NRF) Grant 2021R1A2C2012350 (SK, EL).

\appendix

\section{Saddles from $(\sigma+r,\tau+s)$ parallelograms}

In this appendix, we shall find more parallelogram saddle points,
extending the ideas of section 2.
We shall find saddles which take the form of $(\sigma_r,\tau_s)\equiv(\sigma+r,\tau+s)$-parallelograms for given $\delta_I,\sigma, \tau$ where $r,s \in \mathbb{Z}$. For simplicity, we only consider
the case with upper sign in (\ref{delta-classify}),
with $\sum_I\delta_I -\sigma - \tau = -1$.
At given $\sigma_r,\tau_s$, one can always find unique $n_I^{(r,s)} \in \mathbb{Z}$
so that $\delta_I^{(r,s)}\equiv\delta_I+n_I^{(r,s)}$ belong to one of the
following two cases:
\begin{equation}\label{chem}
\begin{aligned}
&\delta_I^{(r,s)}=\delta_I + n_I^{(r,s)} =  -a_I^{(r,s)} + b_I (\sigma_r+\tau_s)\ , \quad \pm a_I^{(r,s)} \in (0,1), \; b_I \in (0,1)\ , \\
& \delta_1^{(r,s)}+\delta_2^{(r,s)}+\delta_3^{(r,s)} -\sigma_r - \tau_s = - \sum_{I=1}^3 a_I^{(r,s)} = \mp 1\ , \quad \sum_{I=1}^3 b_I = 1\ .
\end{aligned}
\end{equation}
%For simplicity, we will omit all the superscripts ${(r,s)}$ from now on.
%Whenever $\delta_I, a_I$ appear, they should be understood with $(r,s)$ superscripts.
Results in the two cases are related to each other by the transformation
\begin{equation}\label{delta-rs-classify}
  (\delta_I^{(r,s)},\sigma_r,\tau_s)\rightarrow
  (-(\delta_I^{(r,s)})^\ast,-\sigma_r^\ast,-\tau_s^\ast)\ .
\end{equation}

Our ansatz for the large $N$ distribution $u (x,y) = x \sigma_r + y \tau_s$ with uniform density for $0 < x,y < 1$ will satisfy the saddle point equation when
\begin{equation}\label{con1}
0<\textrm{Im} \left(\frac{\sigma_r}{\tau_s}\right) < \textrm{Im} \left(\frac{\delta_I^{(r,s)}}{\tau_s}\right) <
- \textrm{Im} \left(\frac{1}{\tau_s}\right) \quad  \& \quad  0< \textrm{Im} \left(\frac{\delta_I^{(r,s)}}{\sigma_r}\right) < \textrm{Im} \left(\frac{\tau_s-1}{\sigma_r}\right)
\end{equation}
or
\begin{equation}\label{con2}
0<\textrm{Im} \left(\frac{\tau_s}{\sigma_r}\right) < \textrm{Im} \left(\frac{\delta_I^{(r,s)}
}{\sigma_r}\right) < - \textrm{Im} \left(\frac{1}{\sigma_r}\right)
\quad  \& \quad  0< \textrm{Im} \left(\frac{\delta_I^{(r,s)}}{\tau_s}\right)
< \textrm{Im} \left(\frac{\sigma_r-1}{\tau_s}\right)\ ,
\end{equation}
for the case with upper sign in (\ref{chem}).
For the case with lower sign, the saddle point equation is solved
when the condition obtained by applying the transformation (\ref{delta-rs-classify})
to (\ref{con1}), (\ref{con2}) is met.
When one of the above inequalities is satisfied, $(\sigma_r,\tau_s)$-saddle contributes to the large $N$ free energy as following:
\begin{equation}
\log Z (\delta_I,\sigma,\tau)
=-\frac{\pi i N^2 \delta_1^{(r,s)}\delta_2^{(r,s)}\delta_3^{(r,s)}}{\sigma_r\tau_s}
\ .
\end{equation}

One may take $\textrm{Im} \left(\frac{\tau_s}{\sigma_r}\right) \to 0$ limit from either condition of the above. Then, since $0<a_I<1$, both conditions
are trivially satisfied. Also, as all final quantities are regular in this limit, we can safely conclude that at $\textrm{Im} \left(\frac{\tau_s}{\sigma_r}\right)=0$, i.e. $\tau_s = k\sigma_r \; (k\in \mathbb{R})$, the large $N$ saddle point equation is always satisfied.

The condition `(\ref{con1}) or (\ref{con2})' can be merged into
the following set of inequalities:
\begin{equation}\label{con3}
\begin{aligned}
&-\textrm{min}\left[\textrm{Im}(\sigma),\textrm{Im}(\tau) \right] \leq -\textrm{min} \left[ \frac{a_I^{(r,s)} \textrm{Im} (\sigma)}{1-b_I} ,
\frac{(1-a_I^{(r,s)}) \textrm{Im} (\sigma)}{b_I},
\frac{a_I^{(r,s)} \textrm{Im} (\tau)}{b_I},
\frac{(1-a_I^{(r,s)}) \textrm{Im} (\tau)}{1-b_I}\right]\\
&<\textrm{Im} (\sigma) \textrm{Re} (\tau_s) -\textrm{Im} (\tau) \textrm{Re} (\sigma_r) < \textrm{min} \left[ \frac{a_I^{(r,s)} \textrm{Im} (\tau)}{1-b_I} ,
\frac{(1-a_I^{(r,s)}) \textrm{Im} (\tau)}{b_I},
\frac{a_I^{(r,s)} \textrm{Im} (\sigma)}{b_I},
\frac{(1-a_I^{(r,s)}) \textrm{Im} (\sigma)}{1-b_I}\right]\\
&\leq \textrm{min}\left[\textrm{Im}(\tau),\textrm{Im}(\sigma) \right] \ .
\end{aligned}
\end{equation}
These are the conditions to be met in the case with
upper sign in (\ref{chem}). For the case with lower sign,
the condition to be met is obtained by acting
(\ref{delta-rs-classify}) on (\ref{con3}).
Remarkably, the last condition takes the form of (\ref{con3})
with $a_I^{(r,s)}$ replaced by $\tilde{a}_I^{(r,s)}\equiv a_I^{(r,s)}+1$.
$a_I^{(r,s)}$ satisfy $\sum_I a_I^{(r,s)}=1$, while
$\tilde{a}_I^{(r,s)}$ satisfy $\sum_I\tilde{a}_I^{(r,s)}=2$.

We want to find integers $(r,s)$ which meet the above inequalities
at $I=1,2,3$, when $\delta_I$, $\sigma$, $\tau$ are given.
We have employed an algebraic procedure to solve (\ref{con3}) systematically.
From (\ref{con3}) we can get geometric reasonings of our statements below.
We shall only present the results.

\paragraph{i. $\frac{\tau}{\sigma} \in \mathbb{R}$: collinear case}~\\
\vspace*{-1cm}\subparagraph{1) $\sigma=\tau$}~\\
In this case, \eqref{con3} is satisfied iff $r=s$. These correspond to the $(K,r) = (1,r)$ Bethe roots when $\sigma=\tau$.
According to \cite{Aharony:2021zkr},
the stable Euclidean black hole solution exists iff $r=s$ as we found.

\vspace*{-0.5cm}\subparagraph{2) $\tau = \frac{q}{p}\sigma$
($p,q$ are coprime integers)}~\\
In this case, \eqref{con3} is satisfied iff $s=\frac{q}{p}r$. These correspond to the $(K,r) = (1,r)$ Bethe roots when $\tau = \frac{q}{p}\sigma$
\cite{Benini:2020gjh}.
These saddles are labelled by an integer $l$ as $(r,s)=(pl,ql)$.

\vspace*{-0.5cm}\subparagraph{3) $\tau = k\sigma \;
(k \in \mathbb{R}\backslash\mathbb{Q})$}~\\
In this case, there are `infinitely many' choices of $(r,s)$ satisfying \eqref{con3}, but we cannot explicitly write down possible $(r,s)$. They depend on $\delta_I, \textrm{Im}(\sigma),\textrm{Im}(\tau)$. (See the comments at the end of this
appendix for the true meaning of `infinitely many.')

\paragraph{ii. $\frac{\tau}{\sigma} \notin \mathbb{R}$: non-collinear case}~\\
\vspace*{-1cm}\subparagraph{1)
$\textrm{Im}(\tau) = \frac{q}{p}\textrm{Im}(\sigma)$ ($p,q$
are coprime integers)}~\\
In this case, depending on $(\sigma,\tau)$ the solutions may or may not exist.
If there exist solutions, there are infinitely many.
Given one solution $(r_0,s_0)$ satisfying \eqref{con3}, all other solutions
$(r,s)$ are related to $(r_0,s_0)$ by $(r,s)=(r_0+pl,s_0+ql)$
for some integer $l$. (However, not all values of $l$ are allowed in general.)
The case i-2 is the special case when $r_0=s_0=0$ and \eqref{con3} is satisfied for all integers $l$.

\vspace*{-0.5cm}\subparagraph{2) $\textrm{Im}(\tau) = k\textrm{Im}(\sigma) \; (k \in \mathbb{R}\backslash\mathbb{Q})$}~\\
In this case, there exist `infinitely many' choices of $(r,s)$ satisfying \eqref{con3} just as the case i-3. We cannot explicitly write down possible $(r,s)$. They depend on $\delta_I, \textrm{Im}(\sigma), \textrm{Im}(\tau)$.

In summary, when $\frac{\textrm{Im}(\tau)}{\textrm{Im}(\sigma)} \in \mathbb{Q}$, there can be either `infinitely many' saddle points or no saddle points. When $\frac{\textrm{Im}(\tau)}{\textrm{Im}(\sigma)} \notin \mathbb{Q}$, `infinitely many' saddle points exist.

When we have multiple saddle points labelled by $(r,s)$, we should sum over their contributions to the index. One can easily show that their leading large $N$ entropies after the Legendre transformation are all the same. Namely, the real part of the following entropy function does not depend on the integer shifts
$r,s,n_I^{(r,s)}$,
\begin{equation}
S = -\frac{\pi i N^2 \delta_1^{(r,s)}
\delta_2^{(r,s)}\delta_3^{(r,s)}}{\sigma_r\tau_s}
- 2\pi i \delta^{(r,s)}_I Q_I -2\pi i \sigma_r J_1 - 2\pi i \tau_s J_2
+ 2\pi i (n_I^{(r,s)} Q_I + rJ_1+sJ_2)\ ,
\end{equation}
since the dependence on $n_I^{(r,s)}$, $r$, $s$ is collected to a
pure imaginary constant (the last term).

It is known that there are finite numbers of Bethe roots at $K=1$
\cite{Hong:2018viz}, labeled by finitely many independent $r$'s due to the
symmetry $r\sim r+N$. This happens because $N$ eigenvalues are exactly
equal-spaced. It would be interesting to ask if our $r,s$ enjoy similar
symmetries. However, such a property is impossible to study in our large $N$
continuum formalism. Therefore, when we say that we have found `infinitely
many' solutions for $(r,s)$, this might imply finitely
many solutions whose number scales with large $N$.


\begin{thebibliography}{12345}


%\cite{Witten:1998zw}
\bibitem{Witten:1998zw}
E.~Witten,
%``Anti-de Sitter space, thermal phase transition, and confinement in gauge theories,''
Adv. Theor. Math. Phys. \textbf{2}, 505-532 (1998)
doi:10.4310/ATMP.1998.v2.n3.a3
[arXiv:hep-th/9803131 [hep-th]].
%3180 citations counted in INSPIRE as of 21 Jan 2021

%\cite{Sundborg:1999ue}
\bibitem{Sundborg:1999ue}
B.~Sundborg,
%``The Hagedorn transition, deconfinement and N=4 SYM theory,''
Nucl. Phys. B \textbf{573}, 349-363 (2000)
doi:10.1016/S0550-3213(00)00044-4
[arXiv:hep-th/9908001 [hep-th]].
%336 citations counted in INSPIRE as of 21 Jan 2021


%\cite{Aharony:2003sx}
\bibitem{Aharony:2003sx}
  O.~Aharony, J.~Marsano, S.~Minwalla, K.~Papadodimas and M.~Van Raamsdonk,
  %``The Hagedorn - deconfinement phase transition in weakly coupled large N gauge theories,''
  Adv.\ Theor.\ Math.\ Phys.\  {\bf 8}, 603 (2004)
  doi:10.4310/ATMP.2004.v8.n4.a1
  [hep-th/0310285].
  %%CITATION = doi:10.4310/ATMP.2004.v8.n4.a1;%%
  %407 citations counted in INSPIRE as of 12 Jul 2018



%\cite{Gutowski:2004ez}
\bibitem{Gutowski:2004ez}
  J.~B.~Gutowski and H.~S.~Reall,
  %``Supersymmetric AdS(5) black holes,''
  JHEP {\bf 0402}, 006 (2004)
  doi:10.1088/1126-6708/2004/02/006
  [hep-th/0401042].
  %%CITATION = doi:10.1088/1126-6708/2004/02/006;%%
  %173 citations counted in INSPIRE as of 24 May 2018

%\cite{Gutowski:2004yv}
\bibitem{Gutowski:2004yv}
  J.~B.~Gutowski and H.~S.~Reall,
  %``General supersymmetric AdS(5) black holes,''
  JHEP {\bf 0404}, 048 (2004)
  doi:10.1088/1126-6708/2004/04/048
  [hep-th/0401129].
  %%CITATION = doi:10.1088/1126-6708/2004/04/048;%%
  %231 citations counted in INSPIRE as of 24 May 2018

%\cite{Chong:2005da}
\bibitem{Chong:2005da}
  Z.~W.~Chong, M.~Cvetic, H.~Lu and C.~N.~Pope,
  %``Five-dimensional gauged supergravity black holes with independent rotation parameters,''
  Phys.\ Rev.\ D {\bf 72}, 041901 (2005)
  doi:10.1103/PhysRevD.72.041901
  [hep-th/0505112].
  %%CITATION = doi:10.1103/PhysRevD.72.041901;%%
  %103 citations counted in INSPIRE as of 28 Oct 2018

%\cite{Kunduri:2006ek}
\bibitem{Kunduri:2006ek}
  H.~K.~Kunduri, J.~Lucietti and H.~S.~Reall,
  %``Supersymmetric multi-charge AdS(5) black holes,''
  JHEP {\bf 0604}, 036 (2006)
  doi:10.1088/1126-6708/2006/04/036
  [hep-th/0601156].
  %%CITATION = doi:10.1088/1126-6708/2006/04/036;%%
  %104 citations counted in INSPIRE as of 24 May 2018



%\cite{Kinney:2005ej}
\bibitem{Kinney:2005ej}
  J.~Kinney, J.~M.~Maldacena, S.~Minwalla and S.~Raju,
  %``An Index for 4 dimensional super conformal theories,''
  Commun.\ Math.\ Phys.\  {\bf 275}, 209 (2007)
  doi:10.1007/s00220-007-0258-7
  [hep-th/0510251].
  %%CITATION = doi:10.1007/s00220-007-0258-7;%%
  %421 citations counted in INSPIRE as of 12 Jul 2018

%\cite{Romelsberger:2005eg}
\bibitem{Romelsberger:2005eg}
  C.~Romelsberger,
  %``Counting chiral primaries in N = 1, d=4 superconformal field theories,''
  Nucl.\ Phys.\ B {\bf 747}, 329 (2006)
  doi:10.1016/j.nuclphysb.2006.03.037
  [hep-th/0510060].
  %%CITATION = doi:10.1016/j.nuclphysb.2006.03.037;%%
  %263 citations counted in INSPIRE as of 12 Jul 2018


%\cite{Cabo-Bizet:2018ehj}
\bibitem{Cabo-Bizet:2018ehj}
  A.~Cabo-Bizet, D.~Cassani, D.~Martelli and S.~Murthy,
  %``Microscopic origin of the Bekenstein-Hawking entropy of supersymmetric AdS$_{\bf 5}$ black holes,''
  JHEP \textbf{10}, 062 (2019)
doi:10.1007/JHEP10(2019)062
[arXiv:1810.11442 [hep-th]].
  %%CITATION = ARXIV:1810.11442;%%

%\cite{Choi:2018hmj}
\bibitem{Choi:2018hmj}
S.~Choi, J.~Kim, S.~Kim and J.~Nahmgoong,
%``Large AdS black holes from QFT,''
[arXiv:1810.12067 [hep-th]].
%40 citations counted in INSPIRE as of 07 May 2020

%\cite{Benini:2018ywd}
\bibitem{Benini:2018ywd}
  F.~Benini and P.~Milan,
  %``Black holes in 4d $\mathcal{N}=4$ Super-Yang-Mills,''
Phys. Rev. X \textbf{10}, no.2, 021037 (2020)
doi:10.1103/PhysRevX.10.021037
[arXiv:1812.09613 [hep-th]].
  %%CITATION = ARXIV:1812.09613;%%
  %10 citations counted in INSPIRE as of 23 Apr 2019

%\cite{Agarwal:2020zwm}
\bibitem{Agarwal:2020zwm}
P.~Agarwal, S.~Choi, J.~Kim, S.~Kim and J.~Nahmgoong,
%``AdS black holes and finite N indices,''
Phys. Rev. D \textbf{103}, no.12, 126006 (2021)
doi:10.1103/PhysRevD.103.126006
[arXiv:2005.11240 [hep-th]].
%19 citations counted in INSPIRE as of 01 Jul 2021

%\cite{Choi:2021lbk}
\bibitem{Choi:2021lbk}
S.~Choi, S.~Jeong and S.~Kim,
%``The Yang-Mills duals of small AdS black holes,''
[arXiv:2103.01401 [hep-th]].
%4 citations counted in INSPIRE as of 30 Jun 2021


%\cite{GonzalezLezcano:2020yeb}
\bibitem{GonzalezLezcano:2020yeb}
A.~Gonz\'alez Lezcano, J.~Hong, J.~T.~Liu and L.~A.~Pando Zayas,
%``Sub-leading Structures in Superconformal Indices: Subdominant Saddles and Logarithmic Contributions,''
JHEP \textbf{01}, 001 (2021)
doi:10.1007/JHEP01(2021)001
[arXiv:2007.12604 [hep-th]].
%11 citations counted in INSPIRE as of 23 Feb 2021




%\cite{Hanany:2008sb}
\bibitem{Hanany:2008sb}
A.~Hanany, N.~Mekareeya and G.~Torri,
%``The Hilbert Series of Adjoint SQCD,''
Nucl. Phys. B \textbf{825}, 52-97 (2010)
doi:10.1016/j.nuclphysb.2009.09.016
[arXiv:0812.2315 [hep-th]].
%49 citations counted in INSPIRE as of 01 Jul 2021

%\cite{Gang:2011jj}
\bibitem{Gang:2011jj}
D.~Gang, C.~Hwang, S.~Kim and J.~Park,
%``Tests of AdS$_4$/CFT$_3$ correspondence for $\mathcal{N}=2$ chiral-like theory,''
JHEP \textbf{02}, 079 (2012)
doi:10.1007/JHEP02(2012)079
[arXiv:1111.4529 [hep-th]].
%13 citations counted in INSPIRE as of 01 Nov 2021


%\cite{Closset:2017bse}
\bibitem{Closset:2017bse}
C.~Closset, H.~Kim and B.~Willett,
%``$ \mathcal{N} $ = 1 supersymmetric indices and the four-dimensional A-model,''
JHEP \textbf{08}, 090 (2017)
doi:10.1007/JHEP08(2017)090
[arXiv:1707.05774 [hep-th]].
%47 citations counted in INSPIRE as of 08 Jul 2021

%\cite{Benini:2018mlo}
\bibitem{Benini:2018mlo}
F.~Benini and P.~Milan,
%``A Bethe Ansatz type formula for the superconformal index,''
Commun. Math. Phys. \textbf{376}, no.2, 1413-1440 (2020)
doi:10.1007/s00220-019-03679-y
[arXiv:1811.04107 [hep-th]].
%32 citations counted in INSPIRE as of 08 Jul 2021

%\cite{Hong:2018viz}
\bibitem{Hong:2018viz}
J.~Hong and J.~T.~Liu,
%``The topologically twisted index of $ \mathcal{N} $ = 4 super-Yang-Mills on T$^{2} \times S^{2}$ and the elliptic genus,''
JHEP \textbf{07}, 018 (2018)
doi:10.1007/JHEP07(2018)018
[arXiv:1804.04592 [hep-th]].
%27 citations counted in INSPIRE as of 01 Jul 2021





\bibitem{felder}
G.~Felder and A.~Varchenko, Adv. Math. 156, 44 (2000)
%``The elliptic gamma function and SL(3,Z) x Z^3''
[arXiv:math/9907061].

%\cite{Gadde:2020bov}
\bibitem{Gadde:2020bov}
A.~Gadde,
%``Modularity of supersymmetric partition functions,''
JHEP \textbf{12}, 181 (2021)
doi:10.1007/JHEP12(2021)181
[arXiv:2004.13490 [hep-th]].
%9 citations counted in INSPIRE as of 15 Jul 2021



%\cite{Hosseini:2017mds}
\bibitem{Hosseini:2017mds}
  S.~M.~Hosseini, K.~Hristov and A.~Zaffaroni,
  %``An extremization principle for the entropy of rotating BPS black holes in AdS$_{5}$,''
  JHEP {\bf 1707}, 106 (2017)
  doi:10.1007/JHEP07(2017)106
  [arXiv:1705.05383 [hep-th]].
  %%CITATION = doi:10.1007/JHEP07(2017)106;%%
  %8 citations counted in INSPIRE as of 11 Jun 2018



%\cite{Aharony:2021zkr}
\bibitem{Aharony:2021zkr}
O.~Aharony, F.~Benini, O.~Mamroud and P.~Milan,
%``A gravity interpretation for the Bethe Ansatz expansion of the $\mathcal{N}=4$ SYM index,''
Phys. Rev. D \textbf{104}, no.8, 086026 (2021)
doi:10.1103/PhysRevD.104.086026
[arXiv:2104.13932 [hep-th]].
%6 citations counted in INSPIRE as of 03 Sep 2021



%\cite{Grant:2008sk}
\bibitem{Grant:2008sk}
L.~Grant, P.~A.~Grassi, S.~Kim and S.~Minwalla,
%``Comments on 1/16 BPS Quantum States and Classical Configurations,''
JHEP \textbf{05}, 049 (2008)
doi:10.1088/1126-6708/2008/05/049
[arXiv:0803.4183 [hep-th]].
%47 citations counted in INSPIRE as of 07 Sep 2021

%\cite{Berkooz:2006wc}
\bibitem{Berkooz:2006wc}
M.~Berkooz, D.~Reichmann and J.~Simon,
%``A Fermi Surface Model for Large Supersymmetric AdS(5) Black Holes,''
JHEP \textbf{01}, 048 (2007)
doi:10.1088/1126-6708/2007/01/048
[arXiv:hep-th/0604023 [hep-th]].
%56 citations counted in INSPIRE as of 06 Nov 2021

%\cite{Berkooz:2008gc}
\bibitem{Berkooz:2008gc}
M.~Berkooz and D.~Reichmann,
%``Weakly Renormalized Near 1/16 SUSY Fermi Liquid Operators in N=4 SYM,''
JHEP \textbf{10}, 084 (2008)
doi:10.1088/1126-6708/2008/10/084
[arXiv:0807.0559 [hep-th]].
%31 citations counted in INSPIRE as of 06 Nov 2021

%\cite{Choi:2018vbz}
\bibitem{Choi:2018vbz}
S.~Choi, J.~Kim, S.~Kim and J.~Nahmgoong,
%``Comments on deconfinement in AdS/CFT,''
[arXiv:1811.08646 [hep-th]].
%42 citations counted in INSPIRE as of 31 Aug 2021






%\cite{Basu:2010uz}
\bibitem{Basu:2010uz}
P.~Basu, J.~Bhattacharya, S.~Bhattacharyya, R.~Loganayagam, S.~Minwalla and V.~Umesh,
%``Small Hairy Black Holes in Global AdS Spacetime,''
JHEP \textbf{10}, 045 (2010)
doi:10.1007/JHEP10(2010)045
[arXiv:1003.3232 [hep-th]].
%75 citations counted in INSPIRE as of 24 Dec 2020

%\cite{Bhattacharyya:2010yg}
\bibitem{Bhattacharyya:2010yg}
S.~Bhattacharyya, S.~Minwalla and K.~Papadodimas,
%``Small Hairy Black Holes in $AdS_5 x S^5$,''
JHEP \textbf{11}, 035 (2011)
doi:10.1007/JHEP11(2011)035
[arXiv:1005.1287 [hep-th]].
%66 citations counted in INSPIRE as of 24 Dec 2020

%\cite{Markeviciute:2018yal}
\bibitem{Markeviciute:2018yal}
J.~Markeviciute and J.~E.~Santos,
%``Evidence for the existence of a novel class of supersymmetric black holes with AdS$_5\times$S$^5$ asymptotics,''
Class. Quant. Grav. \textbf{36}, no.2, 02LT01 (2019)
doi:10.1088/1361-6382/aaf680
[arXiv:1806.01849 [hep-th]].
%14 citations counted in INSPIRE as of 24 Feb 2021

%\cite{Markeviciute:2018cqs}
\bibitem{Markeviciute:2018cqs}
J.~Markeviciute,
%``Rotating Hairy Black Holes in AdS$_5\times$S$^5$,''
JHEP \textbf{03}, 110 (2019)
doi:10.1007/JHEP03(2019)110
[arXiv:1809.04084 [hep-th]].
%15 citations counted in INSPIRE as of 24 Feb 2021



%\cite{McGreevy:2000cw}
\bibitem{McGreevy:2000cw}
J.~McGreevy, L.~Susskind and N.~Toumbas,
%``Invasion of the giant gravitons from Anti-de Sitter space,''
JHEP \textbf{06}, 008 (2000)
doi:10.1088/1126-6708/2000/06/008
[arXiv:hep-th/0003075 [hep-th]].
%582 citations counted in INSPIRE as of 03 Sep 2021

%\cite{Grisaru:2000zn}
\bibitem{Grisaru:2000zn}
M.~T.~Grisaru, R.~C.~Myers and O.~Tafjord,
%``SUSY and goliath,''
JHEP \textbf{08}, 040 (2000)
doi:10.1088/1126-6708/2000/08/040
[arXiv:hep-th/0008015 [hep-th]].
%349 citations counted in INSPIRE as of 03 Sep 2021


%\cite{Mandal:2006tk}
\bibitem{Mandal:2006tk}
G.~Mandal and N.~V.~Suryanarayana,
%``Counting 1/8-BPS dual-giants,''
JHEP \textbf{03}, 031 (2007)
doi:10.1088/1126-6708/2007/03/031
[arXiv:hep-th/0606088 [hep-th]].
%90 citations counted in INSPIRE as of 15 Nov 2021

%\cite{Henriksson:2019zph}
\bibitem{Henriksson:2019zph}
O.~Henriksson, C.~Hoyos and N.~Jokela,
%``Novel color superconducting phases of $\cal{N}$ = 4 super Yang-Mills at strong coupling,''
JHEP \textbf{09}, 088 (2019)
doi:10.1007/JHEP09(2019)088
[arXiv:1907.01562 [hep-th]].
%7 citations counted in INSPIRE as of 16 Nov 2021



%\cite{Biswas:2006tj}
\bibitem{Biswas:2006tj}
I.~Biswas, D.~Gaiotto, S.~Lahiri and S.~Minwalla,
%``Supersymmetric states of N=4 Yang-Mills from giant gravitons,''
JHEP \textbf{12}, 006 (2007)
doi:10.1088/1126-6708/2007/12/006
[arXiv:hep-th/0606087 [hep-th]].
%104 citations counted in INSPIRE as of 15 Nov 2021



%\cite{Choi:2019zpz}
\bibitem{Choi:2019zpz}
S.~Choi, C.~Hwang and S.~Kim,
%``Quantum vortices, M2-branes and black holes,''
[arXiv:1908.02470 [hep-th]].
%33 citations counted in INSPIRE as of 31 Aug 2021

%\cite{Nian:2019pxj}
\bibitem{Nian:2019pxj}
J.~Nian and L.~A.~Pando Zayas,
%``Microscopic entropy of rotating electrically charged AdS$_{4}$ black holes from field theory localization,''
JHEP \textbf{03}, 081 (2020)
doi:10.1007/JHEP03(2020)081
[arXiv:1909.07943 [hep-th]].
%29 citations counted in INSPIRE as of 15 Nov 2021

%\cite{Choi:2019dfu}
\bibitem{Choi:2019dfu}
S.~Choi and C.~Hwang,
%``Universal 3d Cardy Block and Black Hole Entropy,''
JHEP \textbf{03}, 068 (2020)
doi:10.1007/JHEP03(2020)068
[arXiv:1911.01448 [hep-th]].
%18 citations counted in INSPIRE as of 16 Nov 2021



%\cite{Choi:2019miv}
\bibitem{Choi:2019miv}
S.~Choi and S.~Kim,
%``Large AdS$_6$ black holes from CFT$_5$,''
[arXiv:1904.01164 [hep-th]].
%36 citations counted in INSPIRE as of 31 Aug 2021

%\cite{Crichigno:2020ouj}
\bibitem{Crichigno:2020ouj}
P.~M.~Crichigno and D.~Jain,
%``The 5d Superconformal Index at Large $N$ and Black Holes,''
JHEP \textbf{09}, 124 (2020)
doi:10.1007/JHEP09(2020)124
[arXiv:2005.00550 [hep-th]].
%10 citations counted in INSPIRE as of 15 Nov 2021


%\cite{Nahmgoong:2019hko}
\bibitem{Nahmgoong:2019hko}
J.~Nahmgoong,
%``6d superconformal Cardy formulas,''
JHEP \textbf{02}, 092 (2021)
doi:10.1007/JHEP02(2021)092
[arXiv:1907.12582 [hep-th]].
%23 citations counted in INSPIRE as of 19 Nov 2021

%\cite{Lee:2020rns}
\bibitem{Lee:2020rns}
K.~Lee and J.~Nahmgoong,
%``Cardy Limits of 6d Superconformal Theories,''
JHEP \textbf{05}, 118 (2021)
doi:10.1007/JHEP05(2021)118
[arXiv:2006.10294 [hep-th]].
%6 citations counted in INSPIRE as of 19 Nov 2021


%\cite{Benini:2020gjh}
\bibitem{Benini:2020gjh}
F.~Benini, E.~Colombo, S.~Soltani, A.~Zaffaroni and Z.~Zhang,
%``Superconformal indices at large $N$ and the entropy of AdS$_5$ $\times$ SE$_5$ black holes,''
Class. Quant. Grav. \textbf{37}, no.21, 215021 (2020)
doi:10.1088/1361-6382/abb39b
[arXiv:2005.12308 [hep-th]].
%24 citations counted in INSPIRE as of 18 Nov 2021



%\cite{Cabo-Bizet:2019eaf}
\bibitem{Cabo-Bizet:2019eaf}
  A.~Cabo-Bizet and S.~Murthy,
  %``Supersymmetric phases of 4d $ \mathcal{N} $ = 4 SYM at large $N$,''
  JHEP \textbf{09}, 184 (2020)
doi:10.1007/JHEP09(2020)184
[arXiv:1909.09597 [hep-th]].
  %%CITATION = ARXIV:1909.09597;%%

%\cite{Colombo:2021kbb}
\bibitem{Colombo:2021kbb}
  E.~Colombo,
  %``The large-N limit of 4d superconformal indices for general BPS charges,''
  JHEP \textbf{12}, 013 (2022)
doi:10.1007/JHEP12(2022)013
[arXiv:2110.01911 [hep-th]].
  %%CITATION = ARXIV:2110.01911;%%



\end{thebibliography}
\end{document}